\documentclass[fleqn,10pt]{aux/wlscirep}
\usepackage[utf8]{inputenc}
\usepackage[T1]{fontenc}
\usepackage{graphicx}	
\usepackage{amsmath}	
\usepackage{amssymb}	
\usepackage{longtable}  
\usepackage{ulem}
\usepackage{ragged2e}   
\usepackage{newfloat}
\usepackage{setspace}
\usepackage{xcolor}
\usepackage{tabularx}
\usepackage{xpatch}
\DeclareFloatingEnvironment[name={Extended Data Figure},fileext=lof]{extendfigure}
\DeclareFloatingEnvironment[name={Supplementary Figure},fileext=lof]{suppfigure}
\DeclareFloatingEnvironment[name={Supplementary Table},fileext=lof]{suptable}
\DeclareFloatingEnvironment[name={Extended Data Table},fileext=lof]{extendtable}
\newcommand{\Teff}{\mbox{$T_{\mathrm{eff}}$}}
\newcommand{\HSTeff}{$\Teff=35,800\pm500$\,K}
\newcommand{\logg}{\mbox{$\log{g}$}}
\newcommand{\HSlogg}{$\logg=7.90\pm0.02$}
\newcommand{\hetoh}{\mbox{$\log(\mathrm{He/H})$}}
\newcommand{\HShetoh}{$\hetoh=-1.90\pm0.20$}
\newcommand{\abun}[3]{\mbox{$\log(\mathrm{#1/H})=-#2\pm#3$}}

\newcommand{\Cuabun}{\abun{Cu}{6.46}{0.14}}
\newcommand{\Znabun}{\abun{Zn}{6.24}{0.21}}

\newcommand{\CIVabun}{$\abun{C}{4.81}{0.06}$}
\newcommand{\Cabun}{$\abun{C}{6.04}{0.17}$}
\newcommand{\totalacc}{$\dot{M}_\mathrm{acc, tot}=4.3\times10^8$\,g\,s$^{-1}$}
\newcommand{\NioverFe}{$\mathrm{Ni/Fe}>2.09$}

\newcommand{\Ion}[2]{#1{\,\sc#2}}

\newcommand{\Msun}{\mbox{$M_{\odot}$}}
\newcommand{\Rsun}{\mbox{$R_{\odot}$}}

\newcommand{\period}{$4.399\pm0.026$\,days}
\newcommand{\amplitude}{$0.120\pm0.018$\,\%}

\usepackage{lineno}

\title{Discovery of a second-generation planet candidate accreting onto a white dwarf}

\author[ ]{
Jamie T. Williams$^{1}$,
Boris T. G{\"a}nsicke$^{1,2,3}$,
Nicholas C. Stone$^{4,5}$,
Detlev Koester$^{6}$,
Benjamin D. R. Davies$^{1,3}$,
Christopher Tong$^{7}$,
David J. Wilson$^{8}$,
Snehalata Sahu$^{1}$,
Andrew Swan$^{1}$,
Thomas G. Beatty$^{4}$,
Sergio H. Ramírez$^{1}$,
Tim Cunningham$^{9}$,
\& Knox S. Long$^{10}$}

\affil[1]{Department of Physics, University of Warwick, Coventry, CV4 7AL, UK}
\affil[2]{Instituto de Astrofísica de Canarias, E-38205 La Laguna, Tenerife, Spain}
\affil[3]{Centre for Exoplanets and Habitability, University of Warwick, Gibbet Hill Road, Coventry, CV4 7AL, UK}
\affil[4]{Department of Astronomy, University of Wisconsin, Madison, WI 53706, USA}
\affil[5]{Racah Institute of Physics, The Hebrew University, 91904, Jerusalem, Israel}
\affil[6]{Institut f{\"u}r Theoretische Physik und Astrophysik, University of Kiel, 24098 Kiel, Germany}
\affil[7]{Department of Physics, MIT, Cambridge, MA 02139, USA}
\affil[8]{Laboratory for Atmospheric and Space Physics, University of Colorado, Boulder, CO 80303, USA}
\affil[9]{Center for Astrophysics | Harvard \& Smithsonian, 60 Garden St, Cambridge, MA 02138, USA}
\affil[10]{Space Telescope Science Institute, 3700 San Martin Drive, Baltimore MD 21218, USA}

\affil[*]{E-mail: jamietwilliams.astro@gmail.com}

\begin{abstract}
\textbf{Many white dwarfs accrete the debris of disrupted planetary bodies, detected in the form of metal-enrichment in their atmospheres, dusty or gaseous circumstellar discs, and photometric transits from debris. The composition of accreted debris has been shown to be diverse yet overall closely resembles Solar System bodies. We report the discovery of a white dwarf accreting material that is unlike any Solar System object. The atmosphere of HS\,0209+0832 is strongly enriched in trans-iron elements including zinc, copper, and niobium but depleted in the canonical rock-forming elements silicon and iron. The composition of the accreted object is consistent with a candidate second-generation planet, formed from the stellar material ejected during the giant phase. The key abundance signature which differentiates it from a first-generation planet is the high enrichment of s-process elements. Photospheric helium and the absence of terrestrial rock-forming elements implies the white dwarf is accreting from the escaped atmosphere of a photo-evaporating giant planet candidate. This is further supported by the detection of a sinusoidal photometric period of \period\ at an amplitude of \amplitude, which we attribute to thermal emission phase variability from a planetary day-night cycle or a transiting cometary tail of an evaporating gas giant.}
\end{abstract}

\begin{document}
\flushbottom
\maketitle

\thispagestyle{empty}



\section*{Main}

HS\,0209+0832 is a hot, young white dwarf with an effective temperature of $\Teff\approx35,000\,$K, a cooling age of $t_\mathrm{cool}\approx5\,$Myr, and a hydrogen-dominated atmosphere with a $\simeq1\%$ trace of helium\cite{Jordan1993}. Such a mixed composition is unusual for a white dwarf of this temperature because helium should sink out of the atmosphere within months \cite{Vennes1988,Koester1989}. A far-ultraviolet spectrum obtained in 1999 with the Space Telescope Imaging Spectrograph (STIS) onboard the \textit{Hubble Space Telescope} (\textit{HST}) revealed the presence of carbon, aluminium, silicon, calcium, titanium, nickel, and zinc in the atmosphere\cite{Wolff2000}, as well as $\simeq100$ lines that could not be identified. Whereas the presence of helium and metals indicates ongoing accretion, the nature of that external source remained unexplained\cite{Wolff2000}. 

\begin{figure*}[t]
\centerline{\includegraphics[width=0.95\hsize]{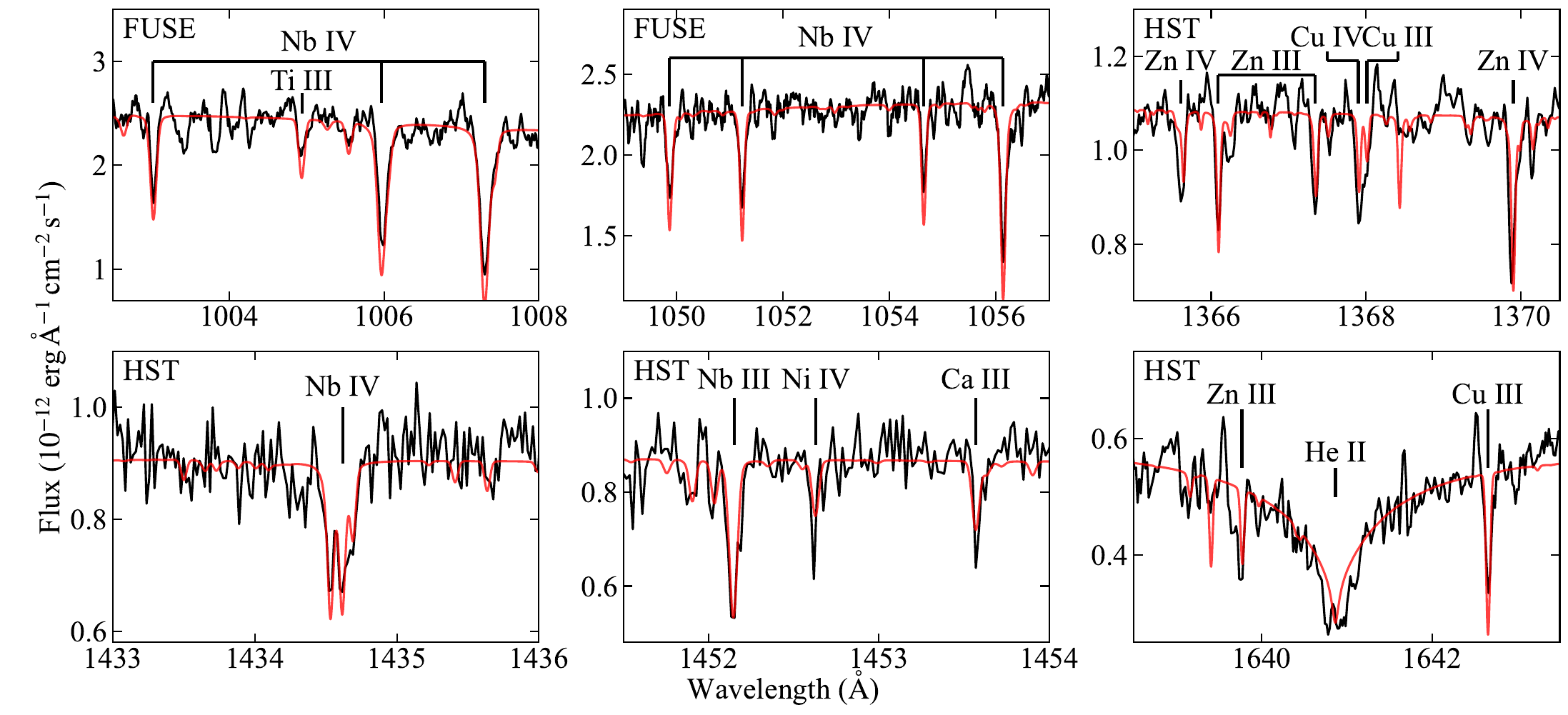}}
\caption{\justifying \textbf{Absorption features in the far-ultraviolet spectrum of HS\,0209+0832}. The \textit{FUSE} and \textit{HST} spectra (both black) of HS\,0209+0832 contain many lines of C, Al, Si, Ca, Ti, Ni, Cu, Zn, and Nb as well as strong hydrogen and helium lines. The unusual structure of the \Ion{He}{II} line is not well reproduced by the model, which could be caused by additional absorption from circumstellar gas. We show the model fit used to obtain metal abundances in red.}
\label{fig:Nb_spec}
\end{figure*}

In the 25 years since that study was published, the origin of photospheric metals detected in many other white dwarfs has been unambiguously confirmed as planetary debris\cite{Zuckerman2007,Gaensicke2012}. This motivated a re-analysis of the observations of HS\,0209+0832 obtained with \textit{HST}/STIS, the \textit{Far Ultraviolet Spectroscopic Explorer} (\textit{FUSE}) and the Ultraviolet and Visual Echelle Spectrograph (UVES) mounted on the Very Large Telescope (VLT), along with the broad-band photometry from PanSTARRS and the \textit{Gaia} parallax (see Methods). In a first step, we determined \HSTeff, a surface gravity of \HSlogg, and a helium abundance of \HShetoh\ (see Extended Data Fig.\,\ref{fig:full_spectrum}
and Extended Data Tab.\,\ref{tab:parameters}
). With these parameters fixed, we computed extensive grids of synthetic white dwarf spectra\cite{koester2010} using detailed lists of metal transitions in the ultraviolet (Extended Data Tab.\,\ref{tab:lines}
). We matched most of the previously unidentified lines to transitions of copper and niobium, with the latter identified using five \Ion{Nb}{III} lines and 57 \Ion{Nb}{IV} lines (see Fig.\,\ref{fig:Nb_spec}). We carefully examined archival \textit{FUSE}\cite{Barstow2014} spectra of 33 known metal-enriched white dwarfs with temperatures $20,000~{\rm K}<\Teff<77,000\,$K and confirm non-detections for niobium within that sample.

We measured the photospheric abundances (see Methods) for nine metals and calculated upper limits on 15 metals (see Extended Data Tab.\,\ref{tab:abundances}
). We excluded any lines originating from the interstellar medium (ISM) in our fit (Extended Data Tab.\,\ref{tab:ISM}
). At the temperature of HS\,0209+0832, radiative levitation can support certain elements against gravitational diffusion. We investigated this effect for all detected elements, as well as for N, P, and S for which we determined upper limits (see Methods). We selected these volatiles because they are commonly detected in hot white dwarfs\cite{Holberg1997,Barstow2014}, thus it is unusual they are not observed in HS\,0209+0832. We find that, for the measured abundances, radiative levitation is negligible for Ca, S, Ti, Ni, Cu, Zn, and Nb. Assuming a steady-state between accretion into, and diffusion out of the white dwarf atmosphere, we computed the diffusion fluxes for each element. We adopt these mass fluxes as being representative of the material accreting onto HS\,0209+0832, as they correct the photospheric abundances for the differential sinking time scales of individual metals (Extended Data Tab.\,\ref{tab:abundances}
). In contrast, the abundances of C, N, Al, Si, and P are subject to radiative support. These elements must still have an external origin\cite{Barstow2003,Koester2014_DAZ}, but our analysis cannot determine their mass fractions within the accreting material. 

\begin{figure*}[t]
\centerline{\includegraphics[width=0.95\hsize]{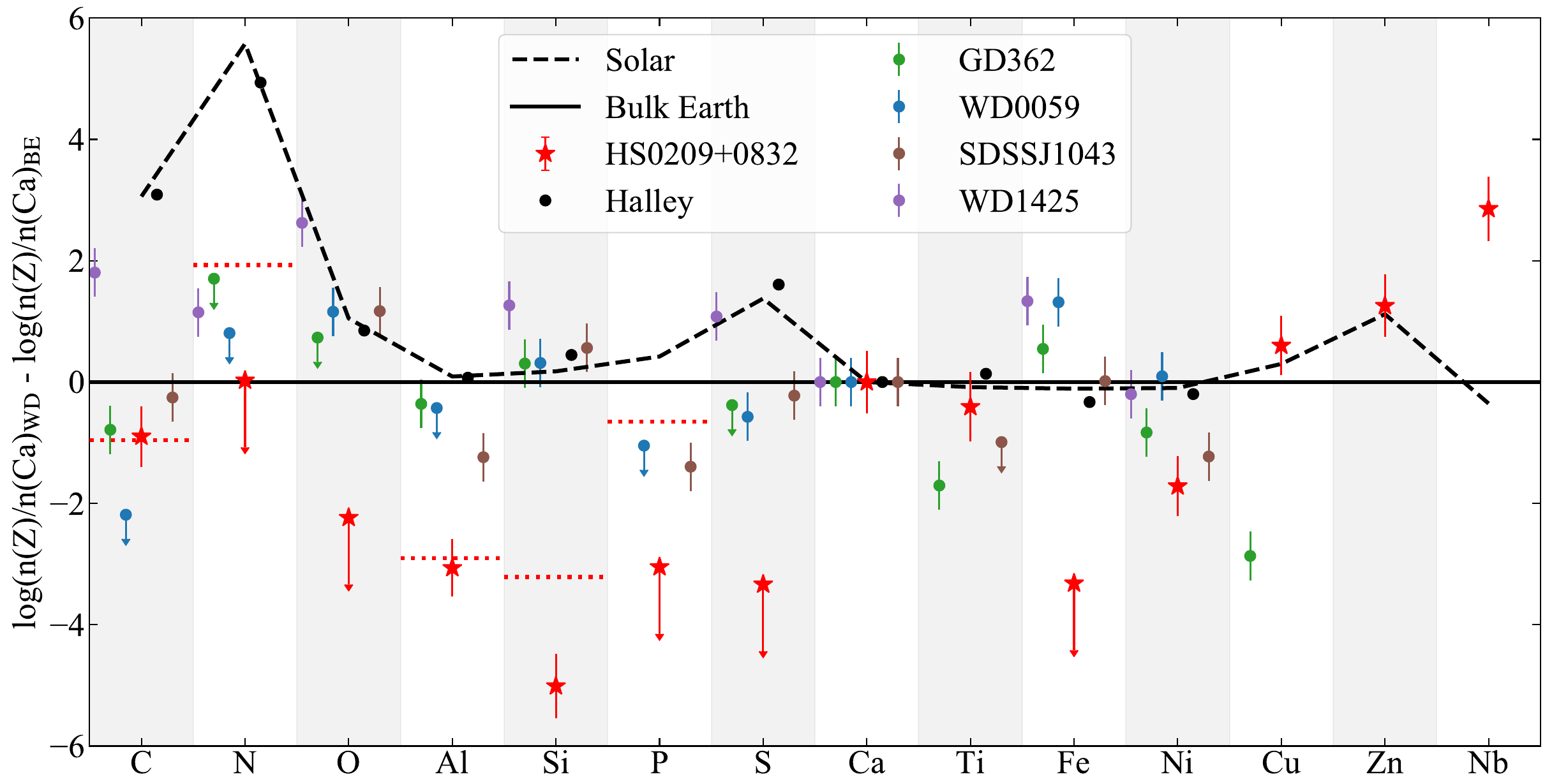}}
\caption{\justifying \textbf{Number abundances of the object accreting onto HS\,0209+0832.} We show the number abundances of HS\,0209+0832 (red stars) with respect to calcium, normalised to bulk Earth\cite{Allegre2001} (black solid line) for all detected metals (Z) and metals with strict upper limits. The error bars represent $1\sigma$ uncertainties and the arrows represent upper limits. The red dashed lines represent the predicted abundance from radiative levitation alone. For comparison, we include the solar abundance pattern\cite{Grevesse2007} (black dashed line), comet Halley\cite{Jessberger1988} (black points), and the abundances of several white dwarfs accreting bulk Earth-like\cite{Zuckerman2007}, core-rich\cite{Williams2025}, mantle-rich\cite{MelisDufour2017}, and icy \cite{Xu2017} material (green, blue, brown, purple points).}
\label{fig:num_abun}
\end{figure*}

We compare the abundances of the parent body accreting onto HS\,0209+0832 to the entire sample of white dwarfs accreting rocky material\cite{Williams2024} and to Solar System meteorites\cite{Nittler2004}, but find no compositional matches (see Fig.\,\ref{fig:num_abun}). Whereas silicon and iron are two major components of planetary bodies in the inner Solar System, with a combined mass fraction of $\simeq48$\% in the bulk Earth\cite{McDonough2000}, we only detect traces of silicon in the atmosphere of HS\,0209+0832, compatible with the support provided by radiative levitation, and no iron. In contrast, the material accreting onto HS\,0209+0832 is rich in nickel, which shares many geochemical properties with iron, and we measure a nickel-to-iron ratio of \NioverFe, compared to $\simeq0.05$ within CI chondrites\cite{Lodders2003} and the bulk Earth\cite{McDonough2000}. We find that the trans-iron elements are strongly enhanced with respect to calcium compared to Solar abundances, in particular niobium which is over three orders of magnitudes more abundant than in the Sun.

Most enriched white dwarfs are depleted in volatile elements such as C, N, and S, but a minority are volatile enriched, which can be due to accreting either icy bodies\cite{Xu2017} or the atmosphere of a giant planet\cite{Gaensicke2019}. 
The only volatile detected in HS\,0209+0832 is carbon (Extended Data Fig.\,\ref{fig:C_lines}
), with a high photospheric abundance which is compatible with fiducial predictions from radiative levitation. However, there are both theoretical and empirical reasons to believe that our fiducial radiative levitation predictions overestimate the importance of this effect. Most importantly, in HS\,0209+0832, the predicted abundances for nitrogen and phosphorus due to radiative levitation are $\simeq2$\,dex higher than the upper limits we derived. As the physics of radiative levitation does not vary from element to element, this implies that carbon must be significantly enhanced within the parent body with respect to other volatile species. In contrast, volatiles such as N, O, P, and S are significantly depleted relative to solar and comet Halley abundances (see Fig.\,\ref{fig:num_abun}), so HS\,0209+0832 cannot be accreting a volatile-rich icy body.

\begin{figure*}[t]
\centerline{\includegraphics[width=0.95\hsize]{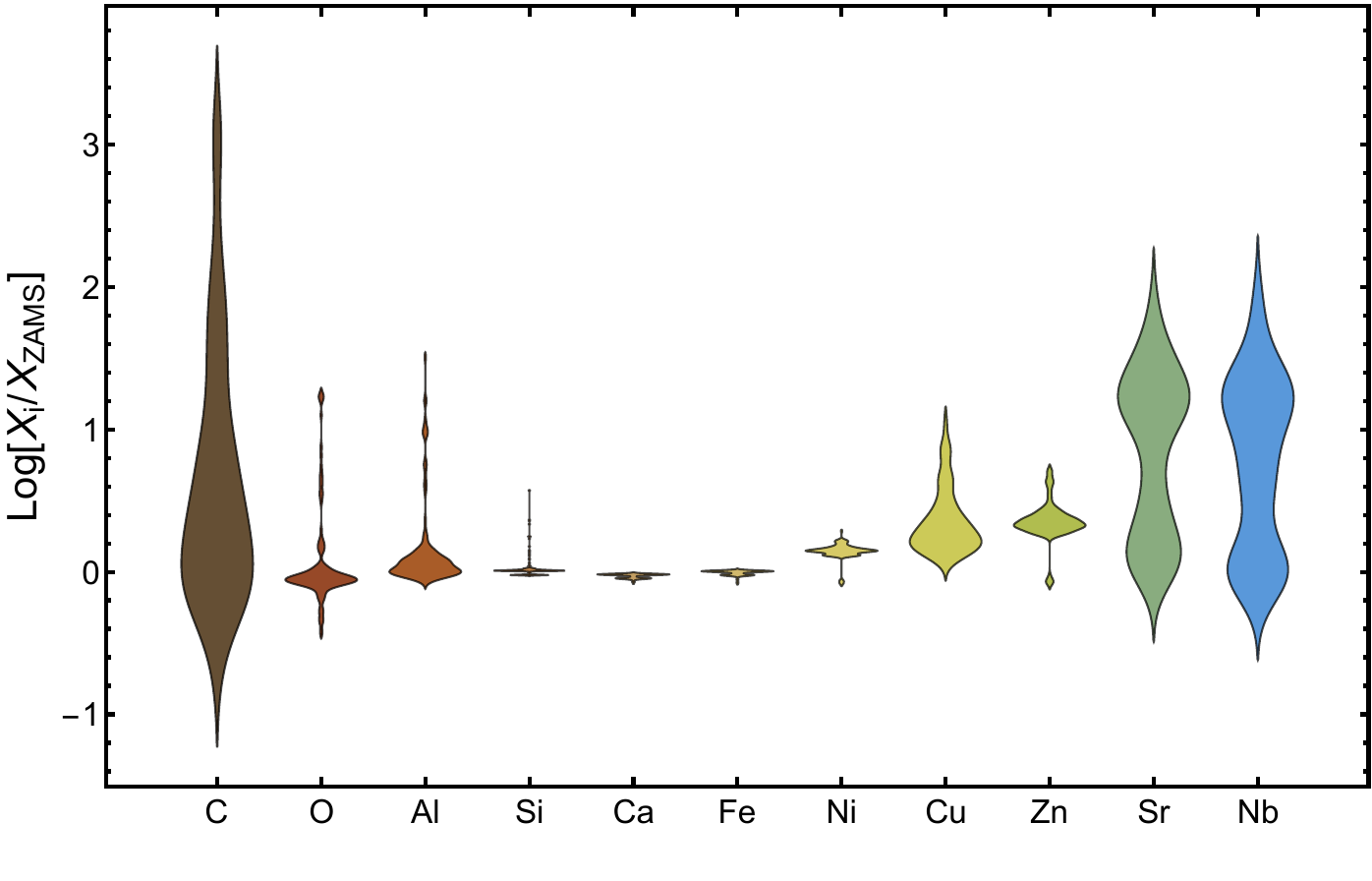}}
\caption{\justifying \textbf{The change in abundances of elements in asymptotic giant branch ejecta.} The width of each metal represents the frequency of the metal having the enhancement relative to the solar abundance\cite{Grevesse2007}, using 138 AGB models. The abundances of oxygen, aluminium, silicon, calcium, and iron share similar abundances to solar abundances (i.e. where $\log [X_\mathrm{i}/X_\mathrm{ZAMS}]=0$) whereas nickel, copper, and zinc have moderate enhancements. There is a wide spread in carbon enhancements, whereas strontium and niobium often have significant enhancements.}
\label{fig:violins}
\end{figure*}

The accretion of helium, the depletion in traditional rock-forming elements, and the enhancement in carbon and trans-iron species implies that the parent body formed in an environment unlike that of a typical proto-planetary disc. In particular, the high carbon and niobium abundances can be explained by the accretion of second-generation planetary material formed within a disc established from a small fraction of the wind-driven mass loss that terminated the life of the asymptotic giant branch (AGB) progenitor of HS\,0209+0832\cite{Perets2010}. The precise abundance patterns in AGB envelope material depend strongly on the progenitor star's initial mass and metallicity\cite{Gallino1998}, but across much of parameter space, AGB envelopes have high carbon abundances. As AGB stars synthesize the overwhelming majority of the Universe's elements formed from slow neutron capture (the ``s-process''), second-generation planets should display unusual abundance patterns featuring strong s-process enhancements, such as a high niobium abundance. This is shown in Fig.\,\ref{fig:violins}, where we examine theoretical predictions from AGB nucleosynthesis models\cite{Buntain2017}. These violin plots show enhancements in AGB winds relative to zero-age main sequence (ZAMS) compositions, and illustrate no enhancements to O, Ca, Si, or Fe. However, we see a modest enhancement in Ni, Cu, and Zn relative to solar and a significant enhancement in niobium. The predicted niobium enhancement as a function of progenitor mass and metallicity is shown in Extended Data Fig.\,\ref{fig:niobium_AGB}
. During the AGB phase, more nickel is produced than iron, which could help explain the large Ni/Fe ratio which we observe in HS\,0209+0832. Whereas the pulsar planets are generally thought to have analogous second-generation origins\cite{PhillipsThorsett1994}, no equivalent has previously been identified orbiting a white dwarf. 

Assuming spherically symmetric mass loss, a second-generation proto-planetary disc is unlikely to form around a single AGB star, but there are several mechanisms which can lead to a disc if a binary companion was present\cite{PeretsKenyon2013}. Evidence for the formation of such a disc has been found at the post-common envelope binary NN\,Ser\cite{Hardy2016} as well as several post-AGB binaries\cite{Kluska2022}.
In particular, a close-in low-mass companion to the progenitor of HS\,0209+0832 would have entered a common envelope event during the AGB phase, resulting in the ejection of the envelope and the formation of a circumstellar disc\cite{Chamandy2025} rich in s-process elements. A second-generation giant planet could have formed within that disc via direct gravitational collapse close to the white dwarf. Alternatively, a first-generation sub-Neptune or super-Earth which was scattered towards the second-generation proto-planetary disk could have accreted a second-generation atmosphere around a first-generation core. The helium detected in the white dwarf atmosphere is probably accreted, which rules out a rocky body fuelling the white dwarf enrichment. A large second-generation disk could form a giant planet on a wider orbit, which was subsequently scattered inwards. However, this would likely require the second-generation planet to form, then a first-generation planet to perturb the second-generation planet towards the white dwarf and for its orbit to circularise, all within 5\,Myrs, the cooling age of the white dwarf\cite{VerasFuller2020}. Known companions which survived common envelope evolution have masses $\gtrsim20\,$M$_\mathrm{Jup}$\cite{Parsons2025}.,This value is above the upper limit of $\gtrsim20\,$M$_\mathrm{Jup}$ of a putative brown dwarf companion to HS\,0209+0832 which we derived from fitting typical, first-generation brown dwarf models\cite{Allard2012,Phillips2020} to the \textit{Spitzer} data. Additionally, a brown dwarf would not accrete enough AGB material to become sufficiently enriched in s-process elements to explain the accretion onto HS\,0209+0832 (see Methods).

\begin{figure*}[t]
\centerline{\includegraphics[width=0.95\hsize]{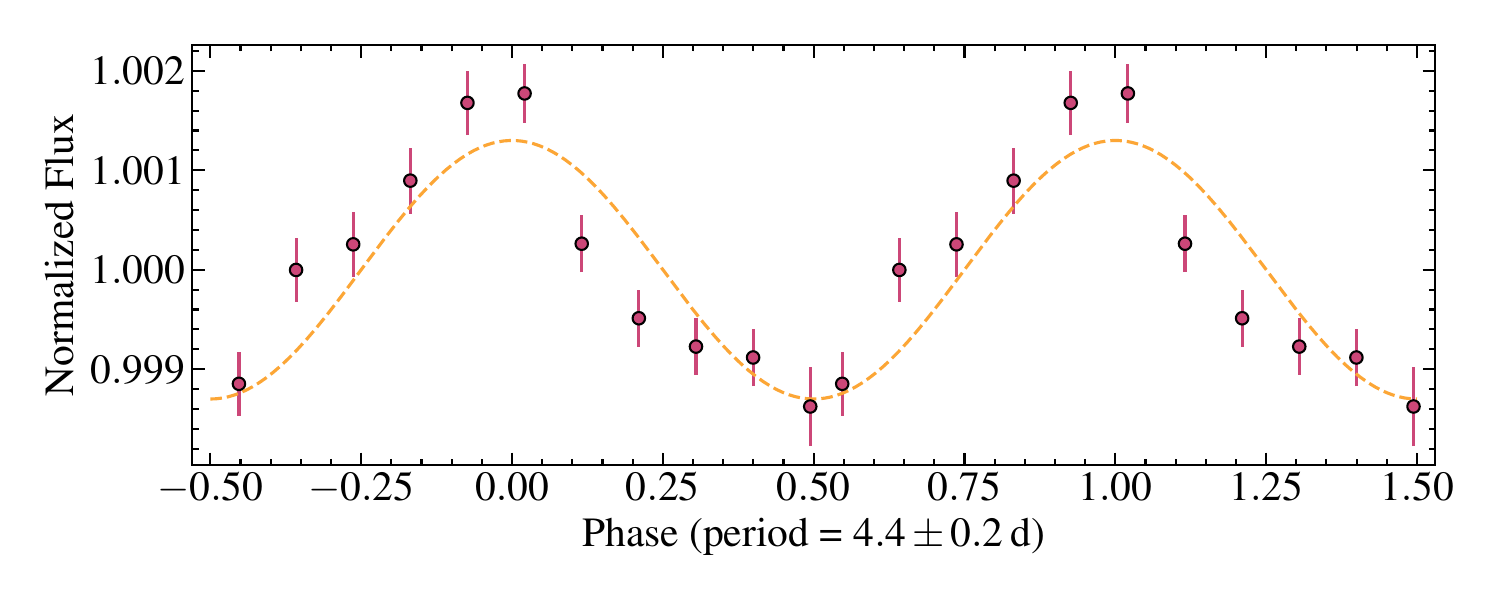}}
\caption{\justifying \textbf{The optical photometric variability of HS\,0209+0832 observed with \textit{TESS}.} The white dwarf was observed in Sectors 42, 43, 70, and 71. We used the 2-minute cadence lightcurves (red points) to identify a sinusoidal signal with a period of \period\ and an amplitude of \amplitude. We show the stitched \textit{TESS} lightcurve binned to 10 hours and folded onto the detected period. The TESS PDCSAP pipeline fluxes and errors were used at all stages. The cycle is repeated for clarity. We include a sinusoid (orange dashed line) with a period and amplitude of the best-fitting parameters.} 
\label{fig:TESS_period}
\end{figure*}

Inspecting the \textit{Transiting Exoplanet Survey Satellite} (\textit{TESS}) observations of HS\,0209+0832 (Extended Data Fig.\,\ref{fig:TPF}
), we detected a sinusoidal signal with a period of \period\ and an amplitude of \amplitude\ (see Fig.\,\ref{fig:TESS_period}, Extended Data Fig.\,\ref{fig:power_spec}
, and Methods). Typical white dwarf spin periods strongly cluster around $\simeq1.25\,$days, with only one out of 20 systems observed with \textit{Kepler} having a rotation period longer than four days\cite{Hermes2017}. The \textit{TESS} photometry can be explained by a giant planet with an orbital separation of 0.04\,AU. Other known close-in white dwarf planets span semimajor axes of $\simeq0.02-0.07\,$au\cite{Gaensicke2019,Vanderburg2020}. The amplitude of the photometric modulation is consistent with the phase curve expected from a young and strongly irradiated phase-locked giant planet (see Methods). The large incident flux of high-energy photons will drive mass loss from this planet\cite{Gaensicke2019,Schreiber2019} which will subsequently accrete onto the white dwarf surface, resulting in the metal-enrichment of HS\,0209+0832. For a planetary radius of 1\,R$_\mathrm{Jup}$, using energy-limited atmospheric escape calculations\cite{Owen2019,Schreiber2019} and the expected EUV and X-ray emission of the white dwarf, we computed an evaporation rate of $\dot{M}_\mathrm{evap}\approx0.11\times10^{13}$\,g\,s$^{-1}$ for a planetary mass of 1\,M$_\mathrm{Jup}$ and $\dot{M}_\mathrm{evap}\approx1.43\times10^{13}$\,g\,s$^{-1}$ for 13\,M$_\mathrm{Jup}$. This is larger than our measured accretion rate onto the white dwarf of $\dot{M}_\mathrm{acc}>4.45\times10^8$. The difference between the evaporation rate and accretion rate is because the white dwarf accretion rate does not include a contribution from hydrogen, helium, and undetected metals. Thermohaline mixing could increase the accretion rate by an order of magnitude but no existing calculations go up to the temperature of HS\,0209+0832\cite{Bauer2018,Dwomoh2023}. Additionally, a substantial proportion of mass is likely ejected from the system by radiation pressure and the dynamical energy injected into the evaporated material from the planetary orbit\cite{Gaensicke2019}.

An alternative explanation to the photometric variability is a cometary tail, similar to that seen in evaporating planets around main sequence stars\cite{Spake2018}. This is further supported by HS\,0209+0832 being spectroscopically variable, with the helium abundance varying by a factor of $2-3$, but over an unknown timescale\cite{Heber1997}. The mass out-flowing from the planet might cross the line-of-sight, explaining the variability in the observed helium abundances, and the unusual structure of the \Ion{He}{II} 1,640\,\AA\ line\cite{Wolff2000}, which is visible in Fig.\,\ref{fig:Nb_spec}.

\clearpage

The abundance patterns in HS\,0209+0832 cannot be explained by direct AGB wind fallback. As niobium comes from the first peak of the s-process, its abundance should be tightly linked to strontium, but the latter is not detected in HS\,0209+0832. However, a class of AGB stars have higher observed niobium abundances relative to strontium than what many models predict\cite{Cseh2022}, of which HS\,0209+0832 could have been a member. 
We also attribute another reason for the unusual Sr/Nb abundance and other compositional anomalies: planetary processing which altered the relative abundances. Several metals observed in HS\,0209+0832 have been observed in giant planet atmospheres such as carbon\cite{Snellen2010}, calcium\cite{Ridden-Harper2016}, titanium\cite{Nugroho2017}, and nickel\cite{Kasper2023}. Aluminium and zinc are predicted to be cloud-forming condensates\cite{Wakeford2017}. Detailed atmospheric modelling of second-generation planets is required to determine likely atmospheric abundances. Despite this complexity, the uniqueness of the observed metals compared to white dwarfs of a similar age\cite{Barstow2014} implies that this is not a typical accretion event. Non-detection of an infrared excess from \textit{Spitzer}\cite{Barber2016} suggests that there is not a substantial fallback accretion disc similar to what is observed in several post-AGB objects\cite{VanWinckel2025}. The abundance pattern in HS\,0209+0832 does not match what is expected from accretion of post-AGB disks. The accretion first proceeds dominated by elements with low condensation temperatures, such as zinc, because the high condensation temperature metals are kept within dust\cite{VanWinckel2025}. Later, dust composed of high condensation temperature metals, such as Nb, accretes. Given that both zinc and niobium are enhanced in HS\,0209+0832, accretion is not dominated by either volatiles or refractory elements. 

Our analysis unambiguously demonstrates that HS\,0209+0832 is accreting material from a second-generation planet candidate. Additional second-generation planet candidates can be identified by observing high carbon abundances and s-process element enhancements, both of which have strong lines in the ultraviolet spectra of hot white dwarfs. Establishing a sizeable sample of such systems will open a window on second-generation planet formation.




\newpage

\section*{Methods}

\subsection*{Spectroscopic observations}

HS\,0209+0832 was observed using \textit{HST}/STIS in the far-ultraviolet (FUV) on 1999 January 10\cite{Wolff2000}. Two exposures of 2,110\,s and 1,140\,s were taken using the MAMA detector and the E140M grating, covering the wavelength range $1,150-1,710\,$\AA\ with a resolving power of $R\simeq45,800$. We used the Hubble Advanced Spectral Products (HASP) spectrum, which combined the individual echelle orders into a single spectrum and co-added the two exposures. The resulting spectrum contains over 200 metal absorption lines, a deep Lyman-alpha line at 1,215\,\AA, and a \Ion{He}{II} line at 1,640\,\AA. There are gaps in the spectrum at $1,416-1,419\,$\AA\ and $1,438.5-1,442\,$\AA\ because this data is flagged as unreliable, which was caused by a large blemish from a shadow on the flat field correction from the repeller wire. The only absorption features present in this region are from metals which have many other lines in the spectrum and hence these gaps do not affect our results.

In addition, two short exposures were taken with the G230LB (600\,s) and G430L (320\,s) gratings, covering the wavelengths 
$1,685-3,065$\,\AA\ and $2,900-5,700\,$\AA\ with resolving powers of $R\simeq700$ and $R\simeq500$. The low resolution of these gratings prevents the detection of metal lines, but we used these data for measuring the white dwarf parameters.

Further FUV spectroscopy of HS\,0209+0832 was taken with \textit{FUSE} on 2002 December 11. The LWRS aperture was used covering a wavelength range of $925-1,180\,$\AA\ with an exposure time of 4,986\,s. The spectra across the four channels were combined, with a resolving power of $R\simeq20,000$. The \textit{FUSE} spectrum contains dozen of metal lines, several hydrogen lines, and two helium lines.

Finally, HS\,0209+0832 was observed with the Ultraviolet and Visual Echelle Spectrograph (UVES) mounted on the Very Large Telescope (VLT)\cite{Napiwotzki2020} on 2000 July $12-15$. Three 900\,s exposures were obtained covering the wavelength range $3,300-6,690$\,\AA\ at an average resolving power of $R\simeq19,155$. There is no data in the region $4,560-4,580$\,\AA\ due to a gap between the detectors. No metal lines are detected in the UVES spectrum, which is unsurprising given the modest quality of that data, and as the transitions of highly ionised metals are much weaker in the optical compared to the FUV. However, we used the hydrogen and helium lines to constrain the white dwarf parameters.

\subsection*{White dwarf parameters}

We computed a finely spaced white dwarf model atmosphere grid\cite{koester2010} spanning effective temperatures of $30,000\leq\Teff\leq40,000\,$K, surface gravities of $7.50\leq\logg\leq8.50$, and a fixed helium abundance of $\hetoh=-2.00$. An initial fit to the Pan-STARRS photometry and \textit{Gaia} parallax (See Extended Data Table\,\ref{tab:parameters}
) resulted in a strong dependence on the assumed reddening, with $\Teff=36,329\pm970\,$K and $\logg=7.918\pm0.024$ for zero reddening and $\Teff=43,075\pm1827\,$K and $\logg=8.049\pm0.030$ for $\mathrm{E}(B-V)=0.01$. Whereas the extinction towards HS\,0209+0832 is low\cite{Lallement2019} ($\mathrm{E}(B-V)<0.01)$, this fit is subject to inherent uncertainties as the photometry only samples the Rayleigh-Jeans tail of the white dwarf spectrum. We also note that the uncertainties of the Pan-STARRS are unrealistically small, resulting in statistically poor fits. 

We therefore adopted a hybrid approach, measuring the effective temperature by fitting the Balmer lines in the VLT/UVES and G430L \textit{HST}/STIS spectra, and measuring the surface gravity from the photometry and parallax. The influence of reddening in  this hybrid method is negligible. Taking the weighted mean of the analysis of the three UVES spectra and one G430L STIS spectrum, with zero reddening, results in $\Teff=35,845\pm535\,$K and $\logg=7.904\pm0.015$. The fit to the ultraviolet spectra with these parameters is excellent (see Extended Data Fig.\,\ref{fig:full_spectrum}
), suggesting that they are well calibrated and that reddening is indeed negligible.

We measured a helium abundance of $\hetoh=-1.90\pm0.20$ from the \Ion{He}{I} (4,026\,\AA, 4,473\,\AA, 4,714\,\AA, 4,923\,\AA, 5,016\,\AA, 5,877\,\AA) and  \Ion{He}{II} (4,686\,\AA) lines in the UVES spectra and the \Ion{He}{II} (1,640\,\AA) line in the E140M STIS spectrum. The uncertainty in \hetoh\ determined from individual lines is 0.2\,dex which limits an assessment of the variability in the He abundance detected in the UVES and STIS spectra.

\subsection*{Photospheric abundances}

With \Teff, \logg, and \hetoh\ fixed to the values determined above (Extended Data Table\,\ref{tab:parameters}
), we generated grids of synthetic spectra for C, Al, Si, Ca, Ti, Ni, Cu, Zn, and Nb with abundances in the range $-10<\log(\mathrm{Z}/\mathrm{H})<-4$ in steps of 0.25\,dex. Unless otherwise stated, the models use atomic data from the Vienna Atomic Line Database \cite{Piskunov1995,Ryabchikova1997,Kupka1999,Kupka2000}.  Interpolating across these model grids we fitted the spectral regions of the \textit{HST}/STIS and \textit{FUSE} spectra which contain absorption lines of a given element using a $\chi^2$ minimization technique (see Extended Data Table\,\ref{tab:lines} 
for the detected absorption features). We detect multiple ISM lines (see Extended Data Table\,\ref{tab:ISM}
) in the spectra, but the radial velocity of these lines differ significantly from the photospheric line velocity, so the former are clearly resolved and can be masked during the fitting procedure. After estimating the abundances for each element in this way, we then computed new model grids allowing one element to vary, and fixing all other elements at the value from the initial analysis, and fitted the \textit{HST}/STIS and \textit{FUSE} data again. This process was repeated until all abundances converged. We report the abundances determined as the weighted mean of the individual fits to each line (or group of lines), and the standard deviation as the uncertainty. 

To determine upper limits\cite{Hollands2020} for N, O, P, S, Cl, V, Cr, Mn, Fe, Co, and Sr, we selected a region that contained a strong line in the model but is undetected in the observed spectrum. We vary the metal abundance by 0.01\,dex, determining the $\chi^2$ each time. The $\chi^2$ values are then converted to a likelihood using Jeffrey's prior and then the cumulative distribution function is computed. We set the upper limit to be the 99\,per cent confidence interval. 

HS\,0209+0832 was the first white dwarf with a detection of photospheric zinc\cite{Wolff2000}. We identified 42 \Ion{Zn}{III} and 40 \Ion{Zn}{IV} lines in the STIS spectrum, and used updated atomic data\cite{Rauch2014} in our analysis. The zinc abundances determined from the different ionisation states agree (\Znabun), but we find the overall abundance to be higher than previously determined\cite{Wolff2000} by $\approx0.9\,$dex. We attribute this difference to improvements in model atmospheres and atomic data.

Whereas several \Ion{Cu}{IV} lines were identified previously\cite{Wolff2000} discrepant atomic data prevented an unambiguous measurement of the copper abundance. Using the updated atomic data\cite{Rauch2020}, we find a copper abundance of \Cuabun\ that provides a good fit to all detected lines, and hence a firm detection of photospheric copper. 

The niobium lines were identified using the National Institute of Standards and Technology (NIST) atomic line database\cite{NIST,Iglesias1955,TauheedReader2005}. All sufficiently strong lines predicted by the synthetic spectra match closely to those in the observed spectrum. We observe no discrepancy between the \Ion{Nb}{III} and \Ion{Nb}{IV} lines. The niobium abundance in the \textit{HST}/STIS and \textit{FUSE} spectra agree within errors.

We note that fitting the \Ion{C}{IV} resonance doublet (1,548 and 1,550\,\AA) results in \CIVabun, i.e. $\simeq1.2\,$dex larger than \Cabun\ determined from the \Ion{C}{II} and \Ion{C}{III} lines, (see Extended Data Fig.\,\ref{fig:C_lines}
). The overlap in wavelength between the \textit{HST}/STIS and FUSE spectra allows independent measurements using the \Ion{C}{III} 1175\,\AA\ multiplet. The abundances determined from the two spectra taken  $\approx4\,$years apart agree within the uncertainties. This situation is reminiscent of GD\,394, a similarly warm ($T_{\mathrm{eff}}=35,700$\,K) white dwarf, in which high ionisation lines of carbon, nitrogen, and phosphorus were detected that were too strong compared to the photospheric abundances measured using low ionisation states\cite{Wilson2019}. Given that the high-ionisation lines in GD\,394 have the same velocity as the lower ionisation photospheric transitions, their origin is most likely a layer of hot gas just above the white dwarf photosphere\cite{Wilson2019}. In HS\,0209+0832, the \Ion{C}{IV} lines also have radial velocities that match the \Ion{C}{II} and \Ion{C}{III} lines, so we argue that the \Ion{C}{IV} lines must originate in gas very close to the photosphere as well. Such a hot ``chromosphere'' is to be expected as the white dwarf is accreting planetary material with relatively low local mass flow rates\cite{WoelkBeuermann1993}, and has been detected in soft X-rays in the debris-enriched white dwarf G\,29-38\cite{Cunningham2022_xray}.

We find a similar discrepancy in the titanium abundances using the \Ion{Ti}{IV} ($\abun{Ti}{5.79}{0.12}$, using the Kurucz line lists\cite{Kurucz1995}) and \Ion{Ti}{III} ($\abun{Ti}{6.34}{0.23}$, using the VALD atomic data) lines. We detect a single \Ion{Ti}{III} line in the \textit{FUSE} spectrum, which agrees with the abundance measured using the \textit{HST}/STIS spectrum. Just as for carbon, the velocities for both ionisation states agree thus we suggest hot gas close to the white dwarf surface as the cause for the higher abundance determined from the higher ionisation lines.

\subsection*{Radiative levitation}

At the effective temperature of HS\,0209+0832, radiative levitation may provide support against gravitational settling for specific elements \cite{Vauclair1979,Vennes1988,Chayer1994,Chayer1995a,Chayer1995b}. According to previous work, among the elements we detect in the photosphere of HS\,0209+0832, C, Si, and Al are likely susceptible to radiative support\cite{Chayer1995b}, whereas Ca and Ni are not\cite{Chayer1994,Chayer1995b}. The radiative support of Ti, Cu, Zn, and Nb in hot white dwarfs has not been assessed so far.

We quantitatively estimated the support provided for each of the detected elements (C, Al, Si, Ti, Ca, Cu, Zn, Nb), and a number of elements with upper limits that are of specific interest (N, P, S, Fe). We used our atmosphere code\cite{koester2010} and followed a documented procedure\cite{Koester2014_DAZ} to compute atmosphere structures and synthetic spectra that include both the effects of atomic diffusion and radiative levitation. In this process, we kept \Teff\ and \logg\ fixed to the values in Extended Data Table\,\ref{tab:parameters}
. In a first step, we computed the equilibrium abundances for each of the above elements, i.e. assuming no external accretion, depositing a small amount of the element within the atmosphere, and iterating the atmosphere structure until the gravity (downward) and radiative levitation (upward) result in a zero velocity of the element under consideration. We found that radiative support is entirely negligible for S, Ca, Ti, Fe, Ni, Cu, Zn, and Nb. The remaining elements are radiatively supported at equilibrium abundances of $\log(\mathrm{C/H})\simeq-6.1$, $\log(\mathrm{N/H})\simeq-6.5$,  $\log(\mathrm{Al/H})\simeq-7.2$,  $\log(\mathrm{Si/H})\simeq-6.5$ and $\log(\mathrm{P/H})\simeq-6.3$. In a final step, we computed the downward diffusion fluxes for the detected elements that are not affected by radiative levitation (Ca, Ti, Ni, Cu, Zn, Nb, see Extended Data Table\,\ref{tab:abundances}
). Because sinking timescales are so short ($\sim$days), we assume that accretion is in the steady state phase, i.e. the downward diffusion flux is equal to the accretion rate.

The primary conclusion of this analysis is that the material accreted by HS\,0209+0832 is genuinely rich in s-process elements. Another important abundance anomaly to note is the very low $\mathrm{Ni/Fe}\gtrsim2$ (by number), compared to $\simeq0.02$ in the Sun, CI chondrites and bulk Earth\cite{Lodders2003,McDonough2000}.

In the case of low accretion rates it is expected that the photospheric abundances increase with time, saturating at or near the equilibrium value\cite{Chayer1995a,Chayer1995b,Koester2014_DAZ}. At higher rates the effect of radiative support becomes increasingly negligible and the photospheric abundances will reflect the steady-state downwards mass flux. We note that for C and Al the photospheric abundances, $\log(\mathrm{C/H})=-6.04$ and $\log(\mathrm{Al/H})=-7.37$, are close to the levels predicted by radiative levitation, $-6.1$ and $-7.2$, respectively. However, we find a photospheric $\log(\mathrm{Si/H})=-8.26$ which is much lower than the radiative equilibrium support of $-6.5$. A key assumption in the models of radiative levitation is that the atmosphere is entirely radiative, i.e. devoid of bulk motion and the associated velocity fields. However, accretion inherently introduces a downward flux of material, and its effect on the atmosphere is currently not well understood. Some models predict large-scale convection due to thermohaline mixing\cite{Deal2013,Bauer2018}, and whereas the physical validity of these models is not fully established\cite{Koester2015}, it is entirely possible that a sufficiently large inflow of material interferes--and possibly weakens or fully suppresses--radiative levitation. In the absence of a full understanding of these complex processes, we proceed with the most conservative assumption, i.e. adopting the radiative support resulting from our atmosphere models as outlined above. 

With the above caveat in mind, we note that radiative support predicts $\log(\mathrm{N/H})=-6.5$, compared to the upper limit on the photospheric abundance of $\log(\mathrm{N/H})<-8.36$. Compared to solar abundances, $\log(\mathrm{C/N})=0.6$, and given that both elements have similarly strong support by radiative levitation, these numbers may suggest that the abundance of carbon is enhanced in the material accreted by HS\,0209+0832. 

At this effective temperature, helium is not radiatively supported and sinks from the atmosphere on timescales of months\cite{Vennes1988,Koester1989}. In fact, HS\,0209+0832 is in the DB gap, a dearth of white dwarfs with atmospheric helium in the range $30,000<\Teff<45,000\,$K\cite{Liebert1986}. We therefore attribute the presence of helium to active accretion.


Our final conclusion is that HS\,0209+0832 accretes material from an external source at a high rate of \totalacc\ with a composition radically different to anything within the Solar System. 

\subsection*{Photometric period}

HS\,0209+0832 (TIC 337075714) was observed with \textit{TESS} in Sectors 42, 43, 70, and 71 (2021 August 20 to 2021 October 12 and 2023 September 20 to 2023 November 11). For each sector, we retrieved the 2-minute cadence \textit{TESS} photometry, processed by the Science Processing Operations Cente (SPOC)\cite{Jenkins2016}, from the Mikulski Archive for Space Telescopes (MAST) using the {\tt lightkurve} package\cite{Lightkurve2018}.

We combined the Pre-search Data Conditioning Simple Aperture Photometry (PDCSAP)\cite{Stumpe2012} from the four \textit{TESS} sectors into a single lightcurve, removed all data that had a quality flag $>0$, and applied a $>5\sigma$ clipping with respect to the median flux. We identified a significant signal with a False Alarm Probability (FAP) of $\simeq4.4\times10^{-16}$ at a period of \period\ with an amplitude of \amplitude\ within the Lomb-Scargle periodogram computed from the combined lightcurve (Extended Data Fig.\,\ref{fig:power_spec}
; Fig.\,\ref{fig:TESS_period}). 

To estimate the uncertainty in the period $P$, we evaluated the $\chi^2$ statistic on a fine grid of trial periods, minimising at each grid point over the remaining free parameters (amplitude $A$, phase $\phi$, and baseline flux $\bar{F}$) by fitting the model $F(t)=A\sin({\frac{2\pi t}{P}+\phi})+\bar{F}$ to the combined lightcurve $F(t)$. The $1\sigma$ confidence interval on the period was taken as the contiguous range of periods for which $\Delta\chi^2 = \chi^2(P) - \chi^2_\mathrm{min} \leq 1$. We estimated the uncertainty in the amplitude similarly, using a grid of trial amplitudes and evaluating $\Delta\chi^2 = \chi^2(A)-\chi^2_\mathrm{min}\leq 1$ over $P$, $\phi$, and $\bar{F}$.


We also performed the same analysis on each sector individually. We identified a significant signal in Sectors 43 and 71 with periods of $4.6123^{+0.0864}_{-0.0816}$\,days (FAP $\simeq1.3\times10^{-6}$) and $4.2963^{+0.0624}_{-0.0600}$\,days (FAP $\simeq8.0\times10^{-15}$), respectively. The amplitudes of the signals detected in Sector 43 and 71 were $0.179^{+0.034}_{-0.035}$\% and $0.237\pm0.038$\%, respectively. Although the amplitude differs between sectors, we note that both of these amplitude values are consistent with the combined lightcurve within $3\sigma$. For Sectors 42 and 70, we did not find any significant periodicity (i.e., ${\rm FAP}<0.01$). Using the best-fit sinusoidal model from the combined lightcurve, we estimate $3\sigma$ upper limits of $A<0.098$\% and $A<0.070$\% for Sectors 42 and 70, respectively. To assess whether these non-detections are consistent with a persistent signal with the period and amplitude from the combined lightcurve, we performed injection-recovery tests for Sectors 42 and 70, injecting sinusoidal models with randomised phases and amplitudes between 0.02\% and 0.5\%. After 1000 trials for each sector we found that, at the amplitude of the signal from the combined lightcurve, we would expect a recovery rate of $\approx40$\% in each case. It is therefore plausible that the signal is present in Sectors 42 and 70 but remains undetected in the periodograms. The scatter of the photometry is substantial compared to the amplitude of the signal (standard deviation of $\sim2.1$\%), and only $\approx6$ cycles are covered per \textit{TESS} sector. Given these limitations, the results of analysing each sector individually are consistent with a persistent $\approx4.4$\,day modulation in the lightcurve of HS\,0209+0832.

Given the large pixel scale of \textit{TESS} ($\simeq21$" per pixel), it is important to consider the possibility of contamination from nearby sources. Examining the Target Pixel Files (TPFs) (Extended Data Fig.\,\ref{fig:TPF}
), we identified four stars of similar brightness within the vicinity of HS\,0209+0832 ($T=14.37$): \textit{Gaia} DR3 2521867602071554432 (object 2 in Extended Data Fig.\,\ref{fig:TPF}
, TIC 337075717, $T=12.88$), \textit{Gaia} DR3 2521867876949459968 (object 3 in Extended Data Fig.\,\ref{fig:TPF}
, TIC 337075708, $T=12.46$), \textit{Gaia} DR3 2521867602071553920 (object 4 in Extended Data Fig.\,\ref{fig:TPF}
, TIC 337075716, $T=13.91$), and \textit{Gaia} DR3 2521867567711816448 (object 5 in Extended Data Fig.\,\ref{fig:TPF}
, not in the TIC, $G=12.20$).  Since these sources do not have lightcurves provided by the SPOC, we extracted the 2-minute cadence lightcurves of these sources from the TPFs for each sector using a custom aperture for each target. For each sector, we then detrended the lightcurves using all single scale and spike cotrending basis vectors, and stitched each sector into a single lightcurve for each target.

Subjecting these lightcurves to the time-series analysis outlined above did not result in the detection of a signal with a period close to that found for HS\,0209+0832. There are significant periods with much lower FAPs than the detected signal on HS\,0209+0832 found at 6.49\,days, 6.27\,days, 5.50\,days, and 6.31\,days for sources 2-5 in Extended Data Fig.\,\ref{fig:TPF} 
respectively. However, these are likely to be spurious signals associated with the P/2 harmonic of the mid-sector data downlink gap\cite{Colman2024}.

We conclude that HS\,0209+0832 exhibits low-amplitude optical variability with a period of \period. 

\subsection*{Interpretation of photometric period}

The photometric variability detected within the \textit{TESS} data must have an origin that is either intrinsic to the white dwarf, or external, i.e. related to a companion. 

Variability in white dwarfs can be caused by pulsations or the combination of rotation and inhomogeneous surface brightness distributions. The latter may be related to the presence of magnetic fields\cite{Hermes2017b}, or variations in atmospheric abundances across their surface\cite{Caiazzo2023}, or a combination of both. Given that helium abundance variations have been reported for HS\,0209+0832\cite{Heber1997}, associating the photometric variability with inhomogeneities in the surface composition seems a plausible explanation. However, the period of the photometric variability in HS\,0209+0832 of $\simeq4.4\,$days is much longer than those of the other white dwarfs exhibiting surface variations in their helium abundance, ranging from tens of minutes\cite{Caiazzo2023,Cheng2025} to at most $\simeq10$\,h\cite{Pereira2005,Moss2025}. In fact, this period is much longer than the typical spin period of white dwarfs of $\simeq30$\,hours, which is well-established from \textit{Kepler} asteroseismology\cite{Hermes2017}. Only one out of 20 stars in the \textit{Kepler} sample had a rotation period above four days. Moreover, a number of white dwarfs showing surface inhomogeneities of helium are extremely massive ($>1\Msun$) and strongly magnetic ($B\gtrsim1\,$MG), so could be the result of binary mergers, whereas HS\,0209+0832 has an average mass ($0.61\Msun$) and no evidence of magnetism ($B\lesssim30\,$kG). 

We now turn to the possibility that the photometric variability is extrinsic to the white dwarf,
noting that the photometric period falls in between those of the two known close-in giant planets orbiting white dwarfs: WD\,1856+534 ($P = 1.4$\,days\cite{Vanderburg2020}) and WD\,J0914+1914 ($P\simeq10$\,days\cite{Gaensicke2019}). No transits are detected in the \textit{TESS} data, which places an upper limit on the orbital inclination of the planet of $i\lesssim88^{\circ}$. A period of $4.40\pm0.02\,$days is equivalent to a planet orbiting at $d=0.044$\,AU, calculated using Kepler's third law. 
The maximum equilibrium temperature $T_\mathrm{p}$ of the planetary dayside would roughly be:

\begin{equation}
    T_\mathrm{p} = \left( \frac{2}{3} \frac{R_\mathrm{WD}^2}{a^2} T_\mathrm{WD}^4 \right) ^{1/4} \approx 1260\,\mathrm{K},
\end{equation}

\noindent assuming zero albedo and no heat redistribution and using values for white dwarf temperature $T_\mathrm{WD}$ and radius $R_\mathrm{WD}$ from Extended Data Table\,\ref{tab:parameters}
. This planetary temperature is on the low end for hot or ultra-hot Jupiters, which cover $T_\mathrm{p}\approx1,400-2,000\,$K\cite{Baxter2020}. Even taking this optimistic $T_\mathrm{p}$ estimate, a possible planet would need to have a radius of approximately 8\,R$_\mathrm{Jup}$ to match the observed variability amplitude due to thermal re-emission of the incoming stellar irradiation. Alternately, if we assume an albedo of $A_\mathrm{g} = 1$, then the possible planet would need to have a radius of 4\,R$_\mathrm{Jup}$ to match the observed photometric variability amplitude. Both of these radii values are (i) well beyond observed radii of even the most extended hot Jupiters \cite{ThorngrenFortney2018}, (ii) exceeds theoretical radii for cooling sequences of young Jovians \cite{Morley+24}, and (iii) is close to filling the Hill sphere ($\approx 10R_{\rm J}$) of a Jovian at $0.044~{\rm AU}$.  While it is possible such a large effective surface could be produced by an outflowing, highly reflective thermal wind, our default interpretation of the observed phase variability is different.

The short cooling age of the WD ($\approx 5~{\rm Myr}$) means that any second-generation planets will still possess large internal reservoirs of heat, and their self-luminosity may create substantial surface temperatures.  For example, using the Sonora Diamondback substellar evolution models\cite{Morley+24}, we estimate that an approximately 10\,M$_{\rm J}$ would have an effective temperature of $T_{\rm eff} \approx 1700~{\rm K}$ and a radius of $R_\mathrm{p}\sim1.75\,M_{\rm J}$ at $t=5~{\rm Myr}$, which in principle is hot and large enough for thermal emission from this possible planet to explain the observed amplitude of phase variability.

However, intrinsic thermal planetary luminosity will fail to produce phase variability: isotropic thermal emission will simply add a constant offset to the WD light curve. Therefore, we propose the hypothesis that the photometric variability may be caused by a young giant planet that has a heavy atmospheric haze layer on its dayside and a relatively clear nightside. The maximum of the photometric variability would then correspond to when we are seeing the hot intrinsic emission of the planet on its nightside, while the variability minimum would occur when the dayside is visible and we see a minimal reflection signal from the optically thick dayside haze later.

As a consistency check on the haze-covered-dayside, clear-nightside hypothesis, we consider the production and destruction timescales of the haze. The EUV environment of this possible planet is orders of magnitude above that of typical hot Jupiters orbiting FGK stars, where photochemical hazes are already routinely invoked to explain muted transmission features and low geometric albedos\cite{Lavvas2017, Kawashima2018, Gao2020}. Scaling from those works, even modest conversion efficiencies of incident EUV photons into haze particles\cite{Zahnle2009} can build dayside aerosol columns of $\tau \gtrsim 1$. on timescales much shorter than the day-to-night advective timescale, $\tau_{\rm adv} \sim R_\mathrm{p}/v_{\rm wind} \sim$ hours-to-days -- assuming $v_{\rm wind} \sim 1$--$10$ km s$^{-1}$\cite{Showman2009, Komacek2016}. The dayside can therefore plausibly form and maintain an optically thick haze layer.

On the nightside, the planet's high intrinsic temperature of 1,700\,K is close to the thermal decomposition temperatures of most photochemical haze precursors and aerosol monomers\cite{Helling2019, He2020}, so once the dayside EUV source is removed the hot atmosphere should rapidly destroy the haze layer. The destruction timescale should be at most comparable to $\tau_{\rm adv}$, so that particles advected from the dayside are destroyed on roughly the same timescale on which they are delivered. A self-consistent microphysical and circulation calculation is beyond the scope of this work. But the dayside EUV-driven production, together with rapid thermochemical destruction on the hot nightside, gives a plausible way to sustain the asymmetric haze distribution our hypothesis requires.

The non-detection of an infrared excess with \textit{Spitzer}\cite{Barber2016} suggests that it was observed during the planet's hazy dayside phase.

We notice that a similar, albeit shorter period ($1.1$\,days) and higher-amplitude ($0.1\%$) modulation is seen in the \textit{TESS} lightcurve of GD\,394\cite{Wilson2020}. That white dwarf has a nearly identical temperature to HS\,0209+0832 ($\Teff=35,700\pm1,500\,$K)\cite{Wilson2019}, but also displays a large-amplitude ($25\%$) modulation at the same 1.1\,days period in the extreme ultraviolet\cite{Dupuis2000}. Magnetic fields and surface inhomogeneities were ruled out by phase-resolved STIS spectroscopy\cite{Wilson2019}. A close-in planet would suffer mass loss due to irradiation by the hot white dwarf. This would form a cloud of evaporated material around the planet, and if this cloud transits GD\,394, it would result in photo-electric absorption of the extreme ultraviolet flux of the white dwarf\cite{Wilson2019}. We note that the same hypothesis could explain both the photometric variability of HS\,0209+0832, as well as the helium line variability (assuming that the atmosphere of the planet contains and loses helium). Line-of-sight helium might also be the cause of the poor fit of the white dwarf model to the \Ion{He}{ii} 1640\,\AA\ line (bottom right panel of Fig.\,\ref{fig:Nb_spec}), for which no plausible explanation was found\cite{Wolff2000}. Helium absorption has been detected in evaporating planets around main sequence stars\cite{Spake2018,Krishnamurthy2025}, which supports the plausibility of the above scenario.

\subsection*{Abundances incompatible with a first-generation planet}

The most unique abundance pattern of HS\,0209+0832 is the enhancements in s-process elements. Strontium and barium, both s-process elements, have been previously detected in high resolution spectra of ultra-hot Jupiter (UHJ) atmospheres such as KELT-9\,b\cite{AzevedoSilva2022}, WASP-121\,b\cite{Borsato2023}, and WASP-189\,b\cite{Prinoth2023}. Although niobium has not yet been detected in UHJs, it is reasonable to assume it is present given the similar nucleosynthetic origins of s-process elements. However, cross-correlation spectroscopy cannot calculate relative elemental abundances, so we cannot infer abundances of s-process elements in first-generation exoplanets. Using main sequence stars as a proxy for first-generation planet abundances, we identified 100 stars from the Hypatia catalogue with abundances of both calcium and niobium\cite{Hinkel2014}. We found the maximum niobium abundance was $\log (\mathrm{Nb/Ca}) =-3.3$, much lower than $\log (\mathrm{Nb/Ca}) =-0.94\pm0.33$ for HS\,0209+0832. There is no known process to generate such a large niobium enhancement in a first-generation planet compared to its main sequence host star.

A first-generation planet could plausibly survive common envelope evolution and its orbit brought close to the white dwarf due to drag forces\cite{Merlov2021}, but for a planet to be brought $\simeq8\,$R$_\odot$ away from the white dwarf, extreme fine tuning is needed\cite{NordhausSpiegel2013} or multiple envelope ejection events\cite{Lagos2021}. Additionally, the timescale where accretion can occur is much longer than the timescale for decay\cite{Yarza2023}. Therefore, a planet which survived common envelope evolution would not be substantially enhanced in s-process elements, as we see in HS\,0209+0832. We can calculate an upper limit on the accretion rate $\dot{M}_\mathrm{Edd}$ assuming that the planet accretes material during the common envelope phase at the Eddington limit:

\begin{equation}
    \dot{M}_\mathrm{Edd} = \frac{4\pi GM_\mathrm{p} m_\mathrm{p}}{\sigma_\mathrm{T}c}
\end{equation}

\noindent where $c$ is the speed of light, $G$ is the gravitational constant, $M_\mathrm{p}$ is the mass of the planet, $m_\mathrm{p}$ is the proton mass, and $\sigma_\mathrm{T}$ is the Thomson cross section for an electron. For an extremely long common envelope phase of 10,000\,yrs, a giant planet would accrete a maximum mass of $2.2\times10^{-5}\,\mathrm{M}_\mathrm{Jup}$.

\subsection*{Predicted second-generation planet composition}

The first intact exoplanets discovered, the pulsar planets \cite{Wolszczan1992}, are likely of second-generation origin \cite{Phinney1993, Margalit2017}.  In contrast to first-generation planets, which form from a standard proto-planetary disc coeval with its main sequence host star, second-generation planets form after the host star has finished its life on the main sequence, and thus in a very different environment. Although the material and mechanisms that formed the pulsar planets are quite different in detail, their example has prompted some theoretical investigation of a different type of second-generation planet: one which forms from AGB wind material that settles into a disk in the vicinity of a newborn white dwarf \cite{Perets2010}.

The details of second-generation planet formation have only received limited investigation.  Most existing theoretical works consider second-generation proto-planetary disks formed when a portion of the AGB wind is captured by a binary companion star \cite{Perets2010, PeretsKenyon2013}, although in principle such a disk could form around a single white dwarf if either (i) a progenitor binary system is converted into a single star by a common envelope episode during AGB evolution, leaving behind a white dwarf and a post-common envelope disk\cite{Chamandy2025}, or (ii) a small fraction of the AGB wind returns hydrodynamically, as is known to occur in other contexts. For example, in core collapse supernova explosions, a fraction of the initially unbound ejecta can become bound later on and fall back to the centre of the system. A variety of hydrodynamical mechanisms have been proposed to achieve this\cite{Wong2014}; one that seems relatively plausible for the white dwarf context is ejecta deceleration by reverse shocks\cite{Chevalier1989}. Specifically, the material ejected in a very late thermal pulse has very high velocities \cite{Kerber2002} and will overtake the remnants of the slow\cite{McDonald2018} AGB superwind, launching a reverse shock into the final mass ejection that can decelerate it and may bind a fraction of it to the white dwarf. It should be emphasized that these second-generation disks are not purely theoretical: infrared excesses have been detected around a large number of post-AGB stars, pointing to the existence of post-AGB disks\cite{VanWinckel2025} where planets could form\cite{Hardy2016}.

Once present, a second-generation proto-planetary disk can in principle create smaller planets by core accretion or larger ones by Toomre instability, as in first-generation disks.  The composition of such second-generation planets will be strongly affected by geochemical and (if present) atmospheric chemical processing, but their bulk compositions will in any case differ dramatically from that of their first-generation counterparts. For example, AGB winds often have inverted C/O ratios, which permits non-standard condensation sequences in which rarer oxygen gets locked up in volatiles while carbon can bind into e.g. carbides\cite{Kuchner2005}. In such an environment, planets formed inside the ice line will be heavily depleted in oxygen, rather than carbon (as is typical for first-generation planets). If this process produces Furthermore, because AGB stars are the Universe's primary locus of s-process nucleosynthesis \cite{Sanders1967, Gallino1998}, objects formed out of AGB wind material will likely be the most s-process enriched planets in the Universe, with corresponding abundance signatures.

An alternative formation mechanism is that a first-generation planet accretes a second-generation envelope, a so-called ``born-again'' planet. For example, a super-Earth or sub-Neptune planet could be scattered towards the second-generation proto-planetary disk and accrete an atmosphere. This would decrease the time needed for planet formation given that the core is already present. However, if the process of enriching a first-generation planet with AGB material works, it would likely be more common to observe accretion of second-generation material onto white dwarfs because most progenitors would eject an s-process-rich envelope, but HS\,0209+0832 is the only example of this to date. Regardless, it would be impossible to determine whether HS\,0209+0832\,b has a first- or second-generation core given the difficulty of measuring interior abundances. Future work on planet formation models within a second-generation proto-planetary disk could deduce whether a first-generation core is required for formation of a second-generation giant planet.


In this paper, we use the results of detailed stellar evolution and neutron capture nucleosynthesis modelling to explore the parameter space of AGB wind compositions.  Our working hypothesis is that second-generation planets will form out of AGB wind material and that their bulk composition will reflect this.  We employ a large grid of stellar evolution models covering a broad range of zero-age main sequence (ZAMS) masses $M_{\rm ZAMS}$ and metallicities $Z_{\rm ZAMS}$.  The models we employ use the Monash stellar evolution code \cite{Frost1996} to explore the synthesis of s-process elements in the He-rich intershell region\cite{Karakas2007}.  Specifically, we examine 138 different stellar evolution models, with $1 M_\odot \le M_{\rm ZAMS} \le 7.5 M_\odot$.  These models cover a discrete set of $Z_{\rm ZAMS}$ values: $Z_{\rm ZAMS}=0.0001$\cite{Lugaro2012}, $Z_{\rm ZAMS}=0.0006$\cite{Karakas2014}, $Z_{\rm ZAMS}=0.001$\cite{Fishlock2014}, $Z_{\rm ZAMS}=0.0028$\cite{Karakas2018}, and $Z_{\rm ZAMS}=0.007$, $Z_{\rm ZAMS}=0.014$, and $Z_{\rm ZAMS}=0.03$\cite{Karakas2007}.  We use these models to build a grid of AGB wind compositions in the space of $\{M_{\rm ZAMS}, Z_{\rm ZAMS}\}$ and linearly interpolate to fill gaps in this space.  

Lower-mass ($1 M_\odot \lesssim M_{\rm ZAMS} \lesssim 4 M_\odot$) stars synthesize s-process elements during the thermally pulsing AGB in a complex way, relying on $\alpha$-capture onto $^{13}$C to provide a neutron source in the intershell.  The buildup of a $^{13}$C pocket by proton capture is thought to occur during third dredge-up episodes, but is highly sensitive to the details of proton mixing from the envelope, creating a major uncertainty in s-process predictions\cite{Buntain2017}.  In practice, the Monash models put a ``partial mixing zone'' (PMZ) in by hand with fixed mass\cite{Karakas2007}.  For many $\{M_{\rm ZAMS}, Z_{\rm ZAMS}\}$ pairs, we have multiple stellar evolution calculations with different parametrized PMZ masses.  When given the choice, we follow the advice of A. Karakas (private communication) and select the closest PMZ mass to $2\times 10^{-3} M_\odot$ for $M_{\rm ZAMS} \le 3 M_\odot$, $1\times 10^{-3} M_\odot$ for $3 M_\odot < M_{\rm ZAMS} \le 4 M_\odot$, $1\times 10^{-4} M_\odot$ for $4 M_\odot < M_{\rm ZAMS} \le 5 M_\odot$, and no PMZ for higher mass progenitors.  At the highest masses, the details of $^{13}$C production become less relevant, as $\alpha$ captures onto $^{22}$Ne provide a stronger neutron source.

We show distributions of AGB wind elemental abundances in Fig.\,\ref{fig:violins}.  In this figure, the y-axis indicates the AGB wind enhancement in a given element relative to the progenitor star's ZAMS abundance of that element.  The width of the distributions comes from mapping our grid of stellar evolution models onto a bivariate probability density function in $\{M_{\rm ZAMS}, Z_{\rm ZAMS}\}$, where ZAMS masses are assumed to follow a Kroupa mass function for $1 \le M_{\rm ZAMS}/M_\odot \le 8$ and ZAMS metallicites are assumed to follow a log-uniform distribution between $10^{-4} \le Z_{\rm ZAMS} \le 0.03$. Figure \ref{fig:violins} shows that white dwarf progenitors experience little change in O, Ca, Si, or Fe abundances during main sequence and post-main sequence evolution. Carbon abundances can be substantially enhanced or depleted, depending on the details of post-main sequence evolution.  Trans-iron elements are generally enhanced relative to ZAMS abundances. Nickel is enhanced by a factor $\approx 2$; zinc by a factor of $\approx 3-5$; copper by a broad range of factors, up to an order of magnitude.  Actual s-process elements, such as strontium (the center of the first ``s-process peak,'' where neutron absorption cross-sections are minimized) and niobium, can be enhanced by up to two orders of magnitude relative to ZAMS abundances. The progenitor mass of HS\,0209+0832 is $M_\mathrm{prog}=1.34^{+0.36}_{-0.26}\,$M$_\odot$, and the extreme niobium enhancement suggests that the progenitor had a low absolute metallicity, $\log Z \leq -2.75$ compared to absolute solar metallicity of $\log Z_\odot \approx1.9$\cite{Asplund2006}, shown in Extended Data Fig.\,\ref{fig:niobium_AGB}
. The abundance of heavier s-process elements such as barium or lead is inversely proportional to the metallicity of the progenitor star, i.e. lead will be most enhanced for metal-poor progenitors. A detection of lead using a higher quality spectrum with improved atomic data would further constrain the progenitor metallicity.


HS\,0209+0832 is remarkably enhanced in the s-process element niobium, with atmospheric abundances ($\log({\rm Nb/H}) = -6.33$) that exceed even the maximum values ($\log({\rm Nb/H}) \sim -7.5$) seen in our model grid, as can be seen in Extended Data Fig.\,\ref{fig:niobium_AGB}
. Beyond the raw level of niobium, the detailed abundance patterns of HS\,0209+0832 do not match theoretical predictions from AGB winds. In particular, niobium abundances are tightly correlated with strontium abundances in all theoretical calculations of s-process nucleosynthesis. Although we include $\approx100$ strong strontium lines in our synthetic far-ultraviolet spectrum, strontium is not detected in the observed spectrum of HS\,0209+0832, and our upper limit on its abundance ($\log({\rm Sr/H}) < -8.86$) is more than two orders of magnitude below the detected niobium abundance. This is in contrast to AGB nucleosynthesis predictions, which essentially always find strontium enhancements above those of niobium: in our set of nucleosynthetic models, AGB wind compositions have $1.2 < \log(\mathrm{Sr/Nb}) < 1.7$. However, an analogous discrepancy is observed in several barium stars, which were enriched in AGB material from their binary companion\cite{Cseh2022}. A sample of barium stars have higher niobium abundances compared to strontium, which may be linked to a late neutron flux moving the first s-process peak to heavier elements such as niobium\cite{Cseh2022}. 
The existence of these barium stars points to a nucleosynthetic pathway not captured by present-day s-process models which may also be able to explain our observed Sr/Nb ratio in HS\,0209+0832.

Other abundance anomalies appear present in HS\,0209+0832 as well.  One, discussed earlier, is the apparent enhancement in carbon, which is indeed expected for a large subset of AGB stellar winds.  Large portions of AGB parameter space produce C/O ratios greater than unity, potentially setting the stage for non-standard condensation sequences that result in strong oxygen depletion and carbon enhancement, the latter of which we have observed.  Other anomalies are harder to explain, for example the unusual discovery of a Ni/Fe ratio greater than unity.  It is important to note that no theoretical work has been performed on the atmospheres of second-generation planets. Therefore, we do not know the typical cloud-forming species of s-process elements.

We only measure upper limits for the s-process elements Y, Zr, Ba, and Pb. However, we note that the atomic data is not complete for these elements due to the lack of transition strengths ($\log\mathrm{gf}$) for several lines and ionisation states, and hence the upper limits are not very constraining. Y, Ba, and Pb only have data for their first ionisation state and Zr only for its second ionisation state. The upper limits are not strict enough to constrain their presence in the accreted material, but future observations or atomic data of higher ionisation states may detect more s-process elements. There are several additional unidentified lines which do not match any other known metal feature, and so improvements in atomic data are required to investigate the presence of additional s-process elements.

\newpage

\section*{Extended Data}

\begin{extendfigure*}[h]
\centerline{\includegraphics[width=0.95\hsize]{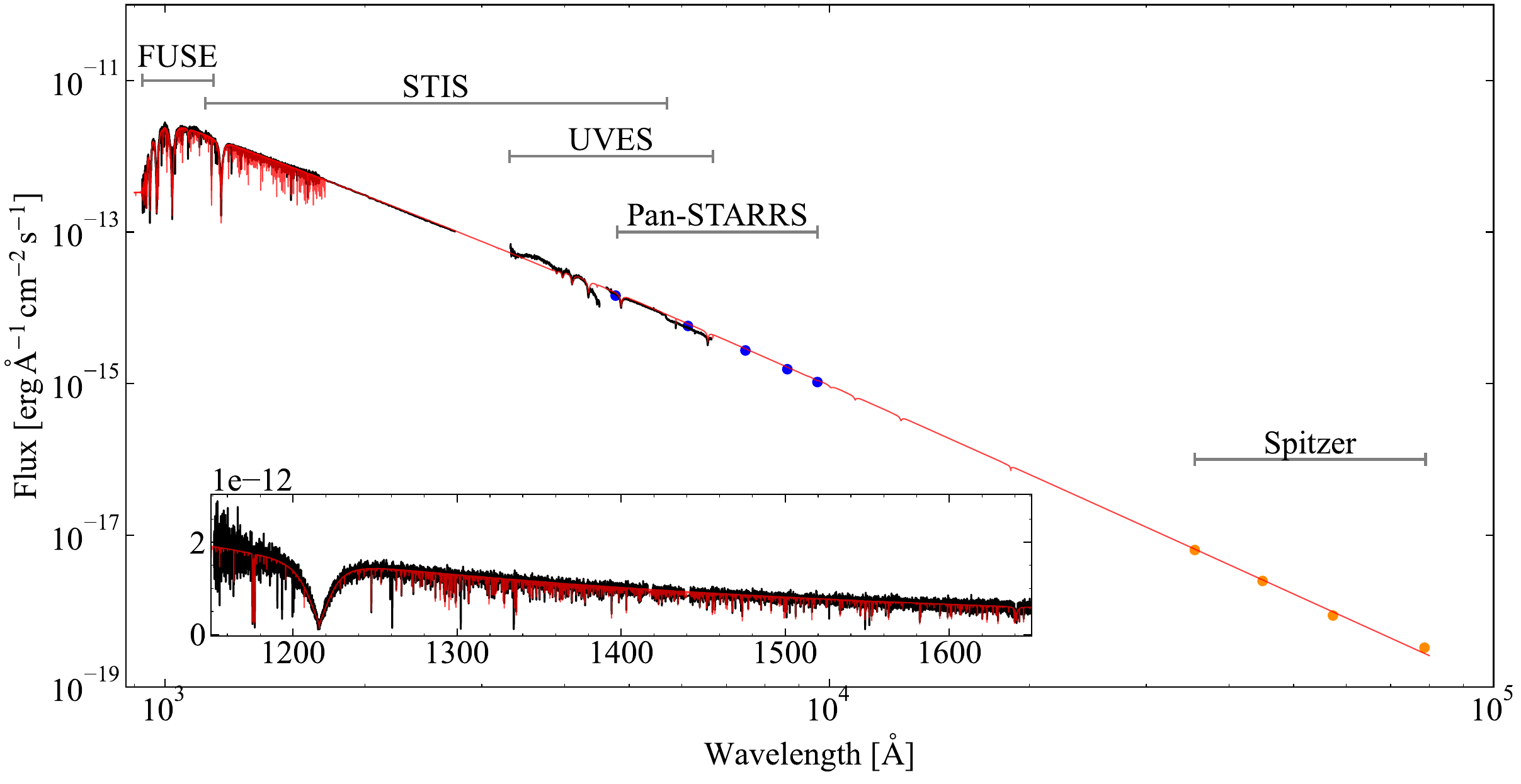}}
\caption{\justifying \textbf{The best model fit to the spectral energy distribution of HS\,0209+0832.} We show the model fit in red and the observed spectra in black. We include the Pan-STARRS and \textit{Spitzer} photometry as blue and orange points. The uncertainty on the photometry are smaller than the symbols. The inset displays the model fit to the \textit{HST}/STIS spectrum.}
\label{fig:full_spectrum}
\end{extendfigure*}

\clearpage
\begin{extendfigure*}[t]
\centerline{\includegraphics[width=0.95\hsize]{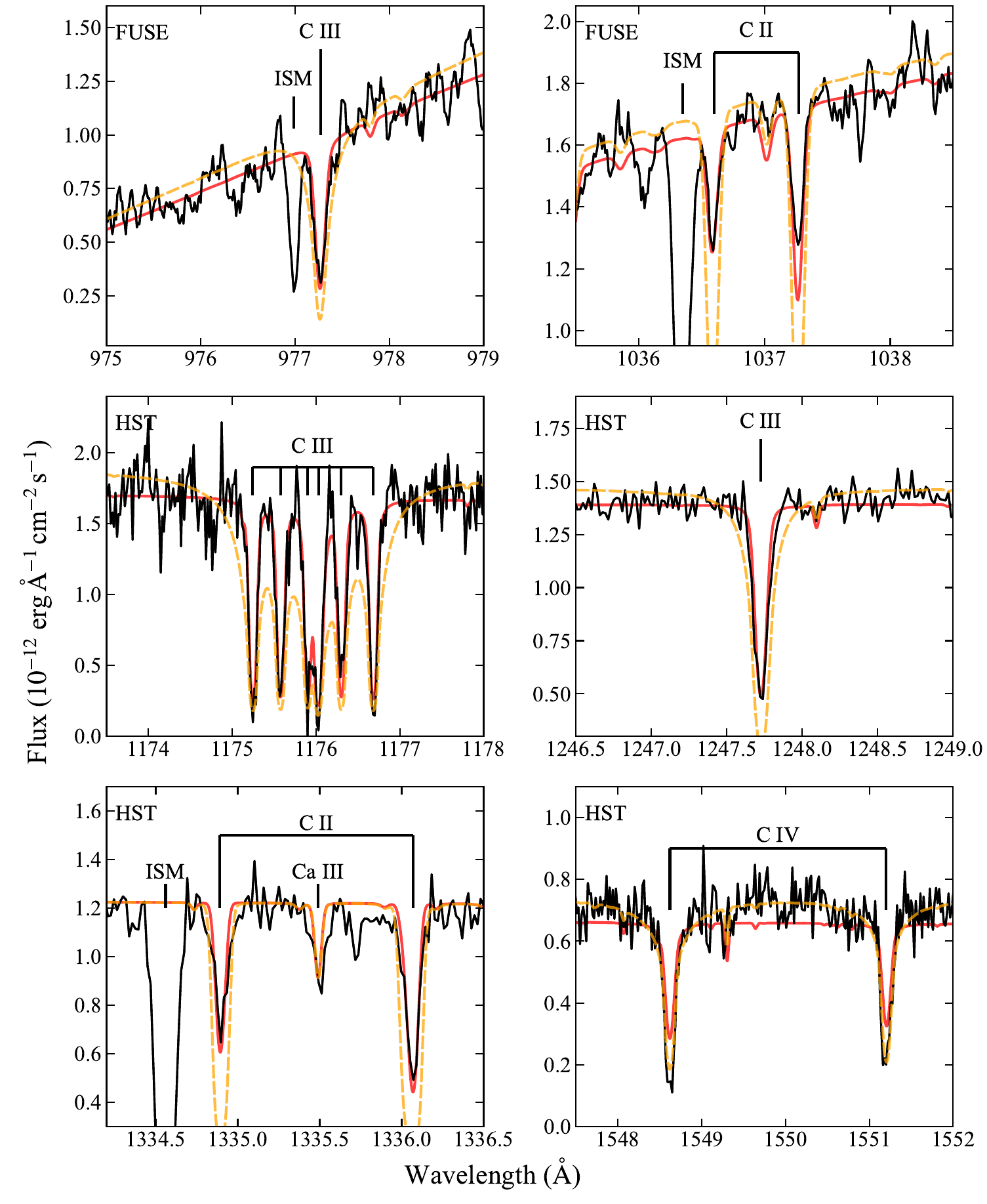}}
\caption{\justifying \textbf{The best model fit to the observed spectrum for carbon lines.}  We show the best fit to the observed spectrum (the top panels are \textit{FUSE} data and the remainder are \textit{HST}/STIS data) in black for \Ion{C}{II} and \Ion{C}{III} lines (\Cabun) with a solid red line and the best fit to the \Ion{C}{IV} lines (\CIVabun) with a dashed orange line. The \textit{FUSE} spectrum has been smoothed with a five-point boxcar.}
\label{fig:C_lines}
\end{extendfigure*}

\clearpage
\begin{extendfigure*}[t]
\centerline{\includegraphics[width=0.95\hsize]{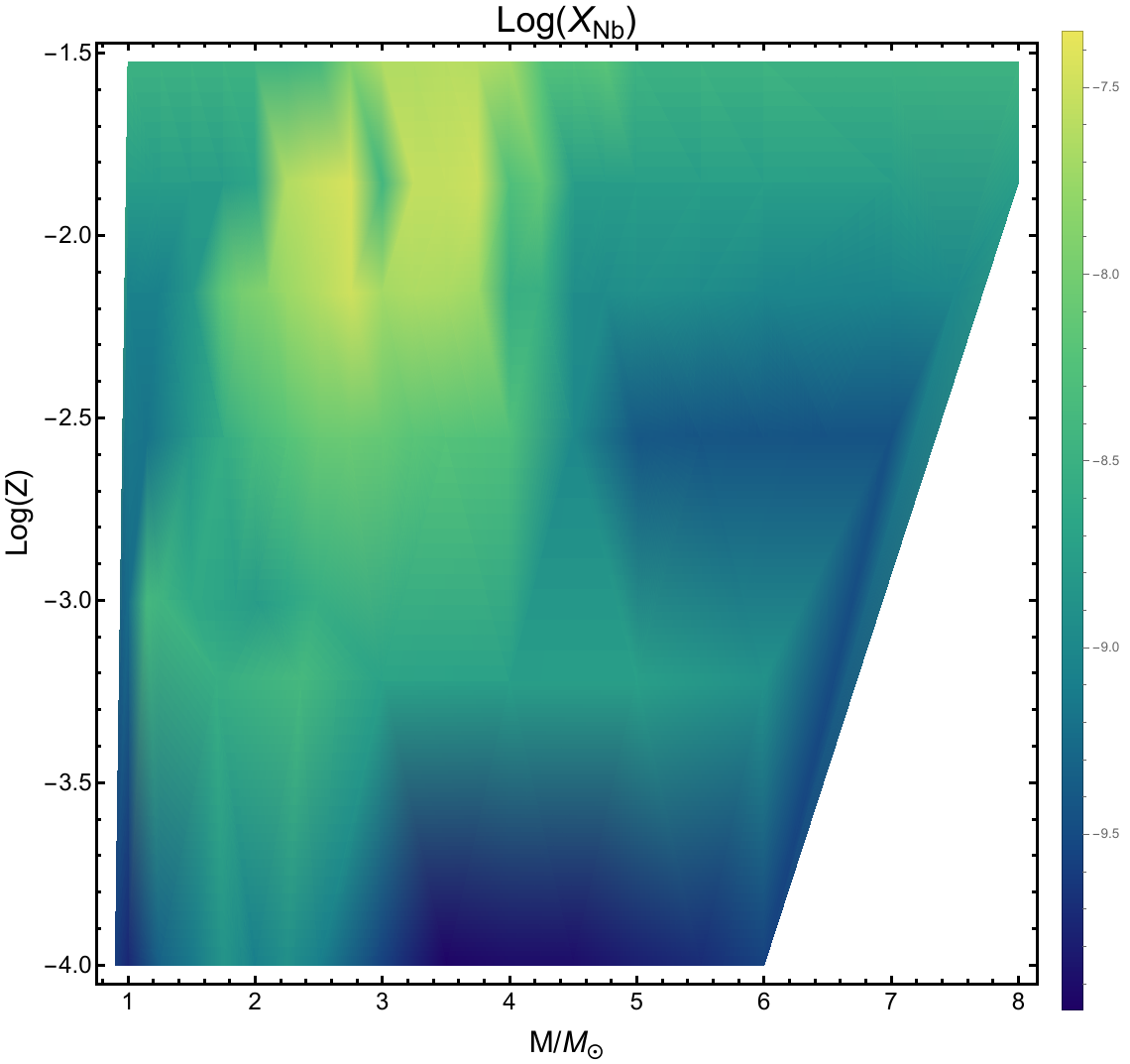}}
\caption{\justifying \textbf{The enhancement of niobium with respect to solar abundances by AGB nucleosynthesis.} Niobium enhancement varies as a function of initial stellar mass, $M_{i}$ and absolute metallicity, $Z_{i}$, with the light regions showing the highest achieved abundances. The highest niobium enhancements are in the parameter range $2\lesssim M_i\lesssim4\,\Msun$ and $-2.5\lesssim Z_i\lesssim-1.5$. The progenitor mass of HS\,0209+0832 is $M_\mathrm{prog}=1.34^{+0.36}_{-0.26}\,\Msun$.}
\label{fig:niobium_AGB}
\end{extendfigure*}

\clearpage
\begin{extendfigure*}[t]
\centerline{\includegraphics[width=0.95\hsize]{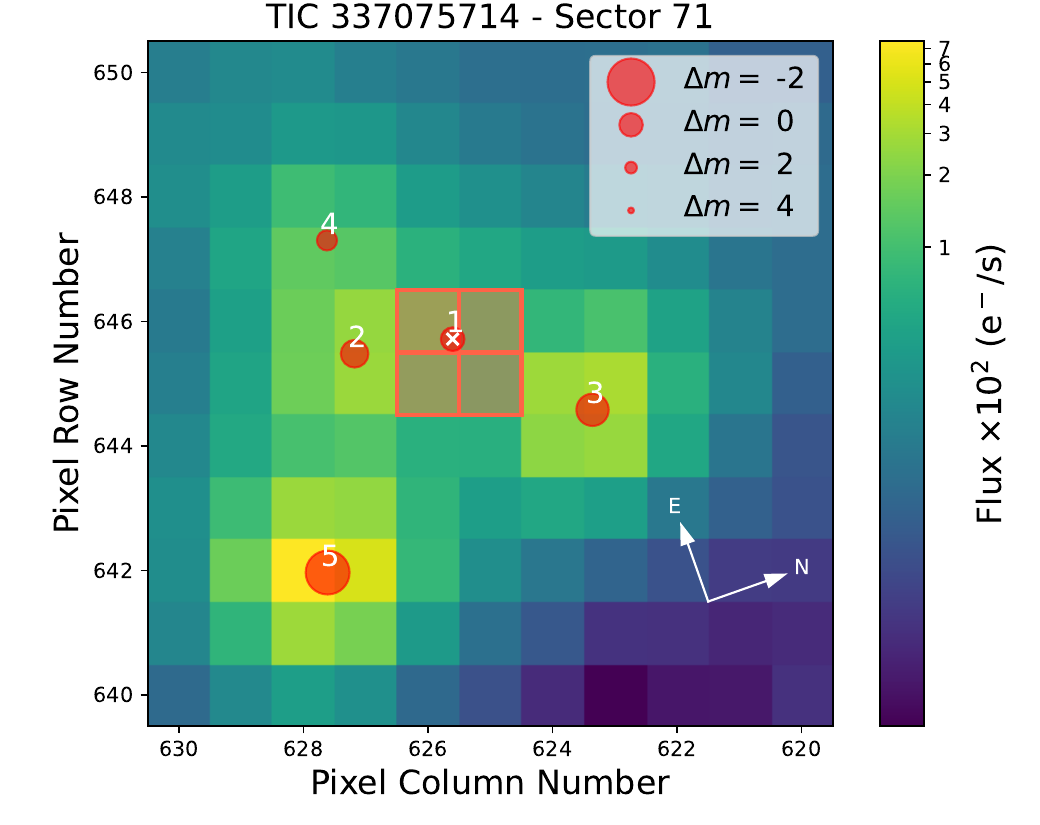}}
\caption{\justifying \textbf{The Target Pixel File (TPF) of HS\,0209+0832 from the \textit{TESS} observations in Sector 71.} The SPOC pipeline aperture is overplotted with shaded red squares. The position of HS\,0209+0832 is demarcated by the number ``1'' and a white cross. All nearby sources within $\Delta m=4$ are displayed as red circles, with their size proportional to their brightness relative to HS\,0209+0832. Each source is labelled by the rank of their projected distances from HS\,0209+0832. The white arrows denote the orientation of the spacecraft during the observations. This figure was created with {\tt tpfplotter}\cite{Aller2020}.}
\label{fig:TPF}
\end{extendfigure*}

\clearpage
\begin{extendfigure*}[t]
\centerline{\includegraphics[width=0.95\hsize]{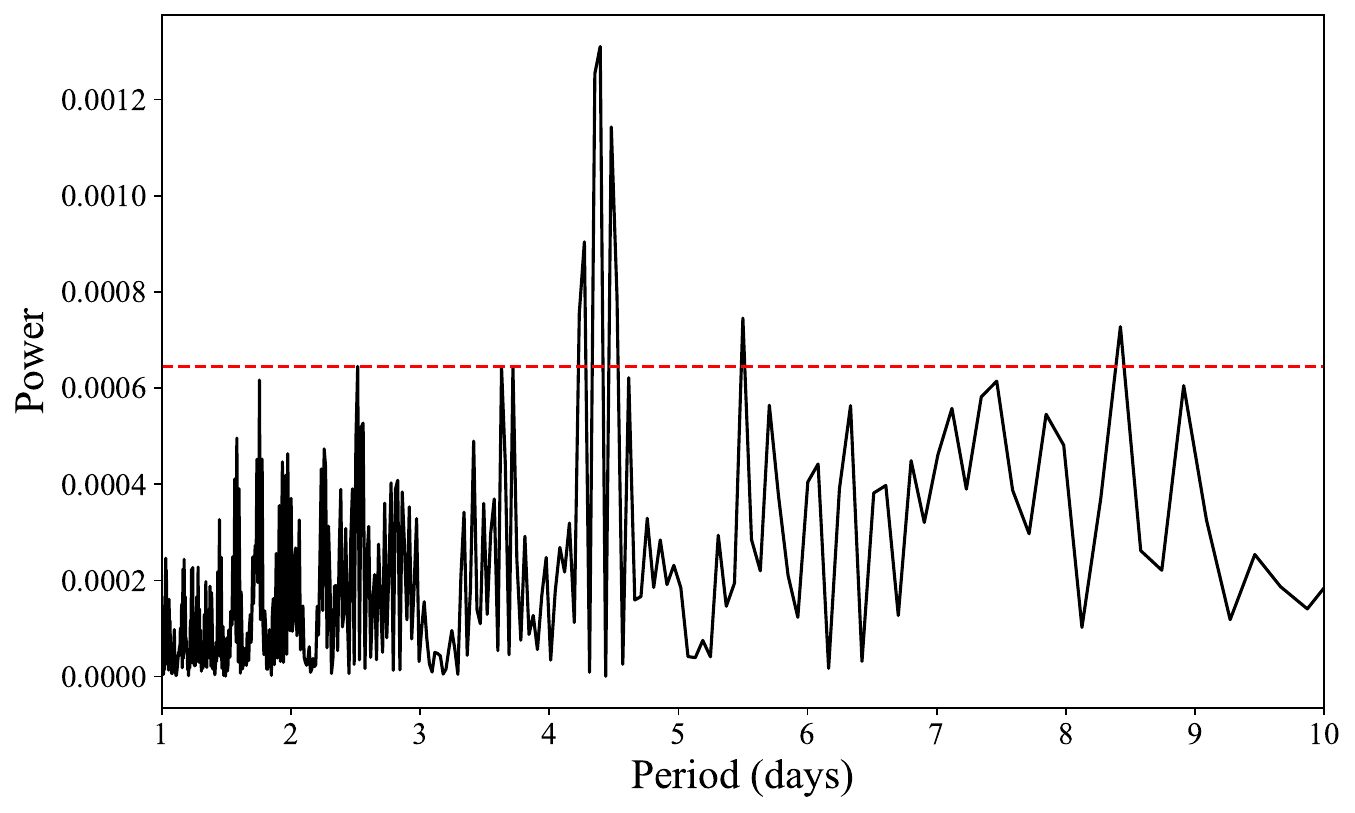}}
\caption{\justifying \textbf{The Lomb-Scargle power spectrum of the combined \textit{TESS} observations obtained in Sectors 42, 43, 70, and 71.} The 0.1\%\,False Alarm Probability level is shown as a horizontal dashed red line.}
\label{fig:power_spec}
\end{extendfigure*}

\clearpage
\begin{extendtable*}
\caption{\justifying \textbf{White dwarf parameters and photometry.} The white dwarf distance is calculated be inverting the \textit{Gaia} parallax. The white dwarf parameters are derived by fitting to spectra and photometry. The progenitor mass and cooling age use the white dwarf parameters. The Pan-STARRS and \textit{Spitzer} data are the photometry used in this study.}
\centering 
\begin{tabular}{c c c}
\hline 
\textit{Gaia} Source ID & 2521867808229983744 & \cite{Gaia2022} \\
parallax [mas] & $12.10\pm0.06$ & \cite{Gaia2022} \\
distance [pc] & $82.6\pm0.3$ & \cite{Gaia2022}, this work \\
\hline
\Teff\ [K]& $35,800\pm500$ & this work \\
\logg\ [dex]& $7.90\pm0.02$ & this work \\
log\,(He/H) [dex] & $-1.90\pm0.20$ & this work \\
mass [\Msun] & $0.61\pm0.01$ & this work \\
radius [\Rsun] & $0.0145\pm0.0002$ & this work \\
progenitor mass [\Msun] & $1.34^{+0.36}_{-0.26}$ & \cite{Cummings2018}, this work \\
cooling age [Myr] & $5.1\pm0.2$ & \cite{Dotter2016,Choi2016,Bedard2020}, this work \\
magnetic field [kG] & $\lesssim30$ & \cite{Hollands2023}, this work \\
\hline
Pan-STARRS \textit{g} [mag] & $13.800 \pm 0.007$ & \cite{Chambers2016} \\
Pan-STARRS \textit{r} [mag] & $14.254 \pm 0.002$ & \cite{Chambers2016} \\
Pan-STARRS \textit{i} [mag] & $14.630 \pm 0.004$ & \cite{Chambers2016} \\
Pan-STARRS \textit{z} [mag] & $14.932 \pm 0.002$ & \cite{Chambers2016} \\
Pan-STARRS \textit{y} [mag] & $15.132 \pm 0.003$ & \cite{Chambers2016} \\
\hline
\textit{Spitzer} $F_{3.6\mu\mathrm{m}}$ [$\mu$Jy] & $262\pm2$ & \cite{Barber2016} \\
\textit{Spitzer} $F_{4.5\mu\mathrm{m}}$ [$\mu$Jy] & $162\pm2$ & \cite{Barber2016} \\
\textit{Spitzer} $F_{5.8\mu\mathrm{m}}$ [$\mu$Jy] & $107\pm8$ & \cite{Barber2016} \\
\textit{Spitzer} $F_{8\mu\mathrm{m}}$ [$\mu$Jy] & $82\pm12$ & \cite{Barber2016} \\
\hline
\label{tab:parameters}
\end{tabular}\\
\end{extendtable*}

\clearpage
\begin{extendtable*}
    \caption{\textbf{Wavelengths of absorption features used to measure metal abundances.} The wavelengths in italics were used to determine upper limits.}
    \centering
    \begin{tabularx}{\textwidth}{rX}
    \hline
        Ion & Vacuum Wavelengths (\AA) \\
        \hline
        \Ion{C}{II} & 1,009.858, 1,010.083, 1,010.371, 1,036.337, 1,037.018, 1,323.951, 1,334.532, 1,335.708 \\
        \Ion{C}{III} & 977.020, 1,174.933, 1,175.263, 1,175.590, 1,175.711, 1,175.987, 1,176.370, 1,247.370 \\
        \Ion{C}{IV} & 1,548.187, 1,550.772 \\
        \Ion{N}{III} & \textit{991.579} \\
        \Ion{O}{III} & \textit{1,153.775} \\
        \Ion{Al}{III} & 1,605.766, 1,611.873 \\
        \Ion{Si}{III} & 1,109.970, 1,206.500 \\
        \Ion{Si}{IV} & 1,122.485, 1,128.340, 1,393.755, 1,402.770 \\
        \Ion{P}{V} & \textit{1,117.977} \\
        \Ion{S}{IV} & \textit{1,062.678} \\
        \Ion{Ca}{III} & 1,278.393, 1,281.554, 1,286.525, 1,298.034, 1,335.128, 1,385.430, 1,453.161, 1,459.793, 1,461.890, 1,463.341, 1,484.868, 1,545.300, 1,555.526, 1,562.464 \\
        \Ion{Ti}{III} & 1,004.669, 1,286.369, 1,289.299, 1,291.625, 1,293.225, 1,294.717, 1,295.884, 1298.697, 1,298.996, 1,327.609, 1,420.439, 1,421.641, 1,421.755, 1,424.136, 1,455.195, 1,498.695, 1,499.177 \\
        \Ion{Ti}{IV} & 1,183.635, 1,195.208, 1,451.736, 1,467.338, 1,469.188 \\
        \Ion{V}{III} & \textit{1,125.699} \\
        \Ion{Cr}{III} & \textit{1,068.391} \\
        \Ion{Mn}{IV} & \textit{1,257.278} \\
        \Ion{Fe}{III} & \textit{1,124.874} \\
        \Ion{Ni}{III} & 1,393.042, 1,396.118, 1,451.499, 1,687.908, 1,692.508 \\
        \Ion{Ni}{IV} & 1,357.064, 1,398.193, 1,399.947, 1,411.451, 1,414.597, 1,421.216, 1,430.190, 1,449.021, 1,452.220, 1,509.101 1,534.710 \\
        \Ion{Co}{IV} & \textit{1,521.644} \\
        \Ion{Cu}{III} & 1,355.442, 1,367.628, 1,376.790, 1,382.538, 1,385.892, 1,430.358, 1,543.450, 1,593.748, 1,600.185, 1,603.138, 1,605.965, 1,606.727, 1,609.752, 1,616.163,  1,626.133, 1,626.405, 1,628.089, 1,628.291, 1,642.202, 1,654.561, 1,681.468, 1,687.119, 1,692.694 \\
        \Ion{Cu}{IV} & 1,309.406, 1,312.715, 1,318.117, 1,340.062, 1,347.785, 1,362.019, 1,367.519, 1,372.113, 1,375.758, 1,380.794, 1,383.120, 1,384.992, 1,388.785, 1,390.416, 1,392.097, 1,410.016, 1,410.531, 1,411.570, 1,412.206, 1,413.016, 1,426.615, 1,429.515, 1,431.928, 1,442.394, 1,443.120, 1,446.299, 1,449.660 \\
        \Ion{Zn}{III} & 1,253.299, 1,262.508, 1,274.384, 1,292.228, 1,303.535, 1,307.350, 1,319.092, 1,323.501, 1,328.367, 1,343.346, 1,359.601, 1,359.799, 1,362.520, 1,364.323, 1,365.706, 1,366.968, 1,373.691, 1,394.910, 1,432.148, 1,456.709, 1,464.180, 1,473.399, 1,490.951, 1,498.778, 1,499.408, 1,500.411, 1,505.913, 1,515.834, 1,552.284, 1,552.947, 1,560.771, 1,581.510, 1,582.036, 1,598.504, 1,600.863, 1,619.601, 1,622.505, 1,629.163, 1,639.320, 1,644.807, 1,651.738, 1,673.077 \\ 
        \Ion{Zn}{IV} & 1,265.707, 1,272.202, 1,272.990, 1,277.080, 1,280.500, 1,283.478, 1,284.711, 1,291.826, 1,292.476, 1,296.652, 1,296.734, 1,301.189, 1,306.657, 1,318.001, 1,320.704, 1,321.215, 1,322.316, 1,322.428, 1,326.774, 1,329.110, 1,329.959, 1,333.326, 1,340.156, 1,342.716, 1,343.750, 1,344.122, 1,347.954, 1,349.876, 1,352.883, 1,357.801, 1,359.477, 1,363.432, 1,363.912, 1,365.253, 1,369.510, 1,375.325, 1,377.615, 1,387.193, 1,387.694, 1,459.944 \\ 
        \Ion{Sr}{IV} & \textit{1,361.154} \\
        \Ion{Y}{II} & \textit{1,339.548} \\
        \Ion{Zr}{III} & \textit{1,356.947} \\
        \Ion{Nb}{III} & 1,451.628, 1,456.692, 1,513.831, 1,524.927, 1,639.512 \\
        \Ion{Nb}{IV} & 981.270, 992.567, 993.538, 1,002.756, 1,005.700, 1,007.015, 1,010.178, 1,013.808, 1,030.271, 1,035.216, 1,044.904, 1,049.610, 1,050.976, 1,054.384, 1,055.874, 1,063.050, 1,065.569, 1,086.748, 1,094.622, 1,103.044, 1,107.844, 1,116.081, 1,119.835, 1,120.243, 1,125.274, 1,127.498, 1,128.611, 1,136.689, 1,316.906, 1,330.601, 1,349.634, 1,363.762, 1,369.762, 1,379.492, 1,386.238, 1,418.884, 1,420.662, 1,424.368, 1,434.140, 1,434.223, 1,444.491, 1,447.491, 1,466.015, 1,472.246, 1,472.664, 1,476.974, 1,487.218, 1,487.260, 1,500.579, 1,508.721, 1,510.832, 1,517.461, 1,524.383, 1,532.606, 1,532.981, 1,534.059 \\ 
        \Ion{Ba}{II} & \textit{1,554.295} \\
        \Ion{Pb}{II} & \textit{1,433.960} \\
        \hline
    \end{tabularx}
    \raggedright
    \label{tab:lines}
\end{extendtable*}

\clearpage
\begin{extendtable*}
\caption{\justifying \textbf{White dwarf abundances.} We include the radial velocity of photospheric lines ($v_\mathrm{r}$), the accretion rate of metals not affected by radiative levitation (RL), and solar abundances\cite{Grevesse2007}.}
\centering 
\begin{tabular}{c c c c c }
\hline 
Z & $\log [\mathrm{Z/H}]$ & $v_\mathrm{r}$ (km\,s$^{-1}$) & $\dot{M}_\mathrm{acc, Z}$ (g\,s$^{-1}$) & solar \\
\hline
C & $-6.04\pm0.17^\mathrm{a}$ & $69.3\pm4.4$ & RL & $-4.60$ \\
N & $<-8.36$ & & & $-5.75$ \\
O & $<-5.32$ & & & $-3.61$ \\
Al & $-7.37\pm0.08$ & $74.5\pm0.2$ & RL & $-5.57$ \\
Si & $-8.26\pm0.24$ & $66.5\pm8.3$ & RL & $-4.49$ \\
P & $<-8.75$ & & & $-6.60$ \\
S & $<-8.26$ & & & $-4.86$ \\
Cl & $<-8.45$ & & & $-6.50$ \\
Ca & $-4.64\pm0.22$ & $78.0\pm2.7$ & $3.8\times10^8$ & $-5.69$ \\
Ti & $-6.24\pm0.33$ & $75.8\pm2.2$ & $8.4\times10^6$ & $-7.10$ \\
V & $<-7.39$ & & & $-8.00$ \\
Cr & $<-7.06$ & & & $-6.36$ \\
Mn & $<-6.48$ & & & $-6.61$ \\
Fe & $<-6.64$ & & & $-4.55$ \\
Co & $<-7.27$ & & & $-7.08$ \\
Ni & $-6.32\pm0.15$ & $76.7\pm1.8$ & $1.1\times10^7$ & $-5.77$ \\
Cu & $-6.46\pm0.14$ & $77.1\pm2.0$ & $9.7\times10^6$ & $-7.79$ \\
Zn & $-6.24\pm0.21$ & $76.4\pm4.1$ & $1.7\times10^7$ & $-7.39$ \\
Sr & $<-8.86$ & &  & $-9.08$ \\
Y & $<-3.48$ & & & $-9.79$ \\
Zr & $<-6.27$ & & & $-9.41$ \\
Nb & $-6.33\pm0.25$ & $68.0\pm7.4$ & $1.9\times10^7$ & $-10.58$ \\
Ba & $<-3.11$ & & & $-9.83$ \\
Pb & $<-5.23$ & & & $-10.00$ \\
\hline
\label{tab:abundances}
\end{tabular}\\
\end{extendtable*}

\clearpage
\begin{extendtable*}
    \centering
    \caption{\textbf{Wavelengths ($\lambda_\mathrm{vac}$) of observed ISM lines and their velocities ($v_\mathrm{ISM}$).}}
    \begin{tabular}{c c c c c c}
        \hline
        \multicolumn{3}{c}{\textit{FUSE}} & \multicolumn{3}{c}{\textit{HST}} \\
        \hline
        Ion & $\lambda_\mathrm{vac}$ (\AA) & $v_\mathrm{ISM}$ (km\,s$^{-1}$) & Ion & $\lambda_\mathrm{vac}$ (\AA) & $v_\mathrm{ISM}$ (km\,s$^{-1}$) \\
        \hline
        \Ion{O}{I} & 936.629 & $2.5\pm3.6$ & \Ion{Si}{II} & 1,190.416 & $4.2\pm0.8$ \\
        \Ion{C}{III} & 977.020 & $-8.4\pm1.5$ & \Ion{Si}{II} & 1,193.290 & $4.7\pm0.6$ \\
        \Ion{O}{I} & 988.773 & $-8.5\pm1.2$ & \Ion{N}{I} & 1,199.550 & $6.6\pm0.5$ \\
        \Ion{N}{III} & 989.799$^\mathrm{a}$ & $18.1\pm2.8$ & \Ion{N}{I} & 1,200.223 & $7.1\pm0.7$ \\
        \Ion{Si}{II} & 989.873$^\mathrm{a}$ & $-4.4\pm2.8$ & \Ion{N}{I} & 1,200.710 & $8.3\pm0.8$ \\
        \Ion{S}{I} & 996.980 & $4.0\pm0.8$ & \Ion{S}{II} & 1,259.519 & $2.0\pm2.3$ \\
        \Ion{C}{II} & 1,036.337 & $1.3\pm0.6$ & \Ion{Si}{II} & 1,260.422 & $5.0\pm0.3$ \\
        \Ion{O}{I} & 1,039.231 & $-0.8\pm1.4$ & \Ion{S}{I} & 1,295.653 & $6.9\pm0.8$ \\
        \Ion{N}{II} & 1,083.990 & $3.6\pm1.9$ & \Ion{O}{I} & 1,302.168 & $7.4\pm0.2$ \\
        \Ion{N}{I} & 1,134.165 & $-2.5\pm2.3$ & \Ion{Si}{II} & 1,304.370 & $4.3\pm0.6$ \\
        \Ion{N}{I} & 1,134.415 & $-6.5\pm1.8$ & \Ion{C}{II} & 1,334.532 & $6.2\pm0.2$ \\
        \Ion{N}{I} & 1,134.980 & $-8.1\pm1.5$ & \Ion{Si}{II} & 1,526.707 & $4.5\pm0.6$ \\
        \hline
    \end{tabular}\\
    $^\mathrm{a}$ Blended.
    \label{tab:ISM}
\end{extendtable*}

\newpage

\section*{Data availability}

All observational data used in this study is available from public data archives. The \textit{HST} and \textit{FUSE} data are available via the \textit{HST} (\href{https://mast.stsci.edu/search/ui/#/hst}{https://mast.stsci.edu/search/ui/\#/hst}) and \textit{FUSE} (\href{https://archive.stsci.edu/fuse/}{https://archive.stsci.edu/fuse/}) MAST archives under program ID 7473 and program ID C026 respectively. The VLT data are available in the ESO archive (\href{https://archive.eso.org/cms.html}{https://archive.eso.org/cms.html}) under program ID 165.H-0588. The Pan-STARRS and \textit{TESS} data are available in the MAST archive (\href{mast.stsci.edu}{mast.stsci.edu}). The \textit{Spitzer} data are available in the \textit{Spitzer} archive (\href{https://irsa.ipac.caltech.edu/data/SPITZER/docs/spitzerdataarchives/}{https://irsa.ipac.caltech.edu/data/SPITZER/docs/spitzerdataarchives/}). The \textit{Gaia} data are available in the \textit{Gaia} archive (\href{https://gea.esac.esa.int/archive/}{https://gea.esac.esa.int/archive/}).

\section*{Code availability}

Specific atmosphere models computed with the white dwarf atmosphere code used in this study\cite{koester2010} are available upon request from the authors.

\section*{Acknowledgements}

We thank for the anonymous referees for their detailed and constructive reports. This research is based on observations made with the NASA/ESA \textit{Hubble Space Telescope} obtained from the Space Telescope Science Institute, which is operated by the Association of Universities for Research in Astronomy, Inc., under NASA contract NAS 5-26555, associated with program 7473. This research used observations made with the \textit{FUSE} observatory under program C026. Based on observations collected at the European Organisation for Astronomical
Research in the Southern Hemisphere under ESO programme 165.H-0588. We acknowledge the use of the National Institute of Standards and Technology (NIST) Atomic Spectra Database. This work made use of \texttt{tpfplotter} by J. Lillo-Box (publicly available in \href{www.github.com/jlillo/tpfplotter}{www.github.com/jlillo/tpfplotter}), which also made use of the python packages \texttt{astropy}, \texttt{lightkurve}, \texttt{matplotlib} and \texttt{numpy}.

\section*{Funding}

This project has received funding from the European Research Council (ERC) under the European Union’s Horizon 2020 research and innovation programme (Grant agreement No. 101020057). BTG acknowledges support from the Fundaci\'on Occident and the Instituto de Astrof\'isica de Canarias under the Visiting Researcher Programme 2022-2025 agreed between both institutions. TC was supported by NASA through the NASA Hubble Fellowship grant HST-HF2-51527.001-A awarded by the Space Telescope Science Institute, which is operated by the Association of Universities for Research in Astronomy, Inc., for NASA, under contract NAS5-26555.

\section*{Author contributions}

All authors contributed to the work presented in this paper. The spectral analysis of HS\,0209+0832 was led by JTW with support from BTG, DK, AS, SS, and SR and the white dwarf atmosphere models used were computed by BTG and DK. KL reduced the \textit{FUSE} observations. JTW, NS, BTG, and CT interpreted the abundances as the signature of a second-generation planet. DJW, BDRD, and TC analysed the \textit{TESS} lightcurves and searched for the periodicity in other photometry. The lightcurve was interpreted by JTW, BTG, NS, and TB. The manuscript was written by JTW with contributions from BTG, NS, and BDRD and additional comments from all authors.

Correspondence and requests for materials should be addressed to Jamie Williams (jamietwilliams.astro@gmail.com).

\section*{Competing Interests}

The authors declare no competing interests.

\newpage

\bibliography{bibliography_HS0209}

\end{document}